\documentclass[aps,prb,twocolumn,showpacs,superscriptaddress]{revtex4}

\usepackage{graphicx}
\usepackage{dcolumn}
\usepackage{bm}

\begin{document}

\newcommand{\bra}[1]{\left\langle#1\right|}
\newcommand{\ket}[1]{\left|#1\right\rangle}
\newcommand{\bracket}[2]{\big\langle#1 \bigm| #2\big\rangle}

\title{
  Dynamical Mean-Field Theory for Molecular Electronics:\\
  Electronic Structure and Transport Properties
}

\author{D. Jacob}
\email{djacob@mpi-halle.de}
\affiliation{Max-Planck-Institute f\"ur Mikrostrukturphysik, Weinberg 2, 06102 Halle, Germany}
\affiliation{European Theoretical Spectroscopy Facility (ETSF)}
\author{K. Haule}
\author{G. Kotliar}
\affiliation{Department of Physics \& Astronomy, Rutgers University, 136 Frelinghuysen Road, Piscataway, New Jersey 08854, USA}

\date{\today}


\begin{abstract}
  We present an approach for calculating the electronic structure and transport
  properties of nanoscopic conductors that takes into account the dynamical correlations
  of strongly interacting $d$- or $f$-electrons by combining density functional theory 
  calculations with the dynamical mean-field theory. While the density functional calculation 
  yields a static mean-field description of the weakly interacting electrons, the dynamical
  mean-field theory explicitly takes into account the dynamical correlations of the strongly 
  interacting $d$- or $f$-electrons of transition metal atoms. As an example we calculate 
  the electronic structure and conductance of Ni nanocontacts between Cu electrodes. 
  We find that the dynamical correlations of the Ni $3d$-electrons give rise to quasi-particle 
  resonances at the Fermi-level in the spectral density. The quasi-particle resonances in turn 
  lead to Fano lineshapes in the conductance characteristics of the nanocontacts similar to those
  measured in recent experiments of magnetic nanocontacts.
\end{abstract}

\maketitle

\section{ Introduction}
\label{sec:introduction}

State of the art for calculating the conductance and current through atomic- and molecular-size
conductors consists in combining {\it ab initio} electronic structure calculations on the level of 
density-functional theory (DFT) with the Landauer formalism or non-equilibrium Green's function
technique \cite{Palacios:prb:01,Taylor:prb:01A}. This methodology has been quite successful for the 
theoretical description of e.g. metallic nanocontacts \cite{Agrait:pr:03} predicting zero-bias 
conductances that are in general in good agreement with experiments 
even in the case of nanocontacts made from magnetic transition metals 
\cite{Viret:prb:02,Untiedt:prb:04,Keane:apl:06,Bolotin:nl:06,Jacob:prb:05,Smogunov:prb:06}.

Recently, however, nanocontacts made from Fe, Co, or Ni have been reported to display Kondo effect
\cite{Kondo:ptp:64,Calvo:nature:09}. This has been infered from the observation of Fano-lineshapes 
\cite{Fano:pr:61} in the low-voltage conductance characteristics similar to those observed in recent 
STM experiments with magnetic adatoms on metal surfaces \cite{Madhavan:science:98,Vitali:prl:08,Neel:prl:08,Neel:prl:07}. 
The observation of the Kondo effect in Fe, Co, and Ni nanocontacts is rather surprising since these materials 
are strong ferromagnets as bulk materials. The Kondo effect, however, is usually at odds with ferromagnetism 
as it results from the screening of a local magnetic moment by antiferromagnetic coupling to the conduction electrons 
\cite{Kondo:ptp:64,Hewson:book}. 

DFT based transport calculations of nanoscopic conductors cannot capture truly many-body 
effects as the Kondo effect that originate from the dynamic correlations of strongly interacting
electrons. Therefore it is necessary to extend the existing methodology in order to account for 
dynamic electron correlations in transport experiments of nanoscopic conductors. 
Recent approaches to include dynamic electron correlations in the {\it ab initio} description 
of quantum transport are based on the GW approximation (GWA) \cite{Thygesen:jcp:07} or the three-body
scattering formalism (3BS) \cite{Ferretti:prl:05}. 
While the GWA is only suitable for weakly correlated systems due to the perturbative treatment of the 
electron-electron interactions, the 3BS is in principle capable of describing more strongly correlated 
systems as it goes beyond perturbation theory.
However, the 3BS does not provide a satisfactory solution of the Anderson impurity problem since the local 
correlations are not taken into account properly. 
More recently, {\it ab initio} electronic structure methods on the level of DFT or the GW approximatuion
have been combined with more sophisticated many-body techniques such as the One-Crossing 
Approximation (OCA) or the Numerical Renormalization Group (NRG) 
in order to account for the Kondo effect in nanoscopic systems containing single magnetic atoms.
\cite{Jacob:prl:09,Lucignano:natmat:09,DiasdaSilva:prb:09,Jacob:prb:10}

In this paper we develop an approach inspired by the success of the Dynamical Mean-Field Theory
(DMFT) in the treatment of correlated solids\cite{Georges:rmp:96,Vollhardt:zpb:97}, to tackle 
the challenges of molecular electronics. DMFT is an approach based on the locality of the self-energy 
and does not require Bloch periodicity. It has been applied before to strongly spatially inhomogeneous 
systems such as alloys near an Anderson transition\cite{Dobrosavljevic:philtrans:98}, 
surfaces\cite{Potthoff:prb:99} and interfaces\cite{Okamoto:nature:04}, multilayered 
heterostructures\cite{Freericks:book:06} and cold atoms in a trap\cite{Gorelik:10}. 
Notice, however, that so far all these studies were restricted to {\it model Hamiltonians} as 
for example the Hubbard model. 

In order to incorporate realistic aspects of the electronic structure we extend the DFT+DMFT 
philosophy\cite{Kotliar:rmp:06,Held:advphys:07} to the case of nanoscopic conductors. Our {\it Molecular} 
DMFT approach takes into account the local dynamical correlations of the strongly interacting 
$3d$- or $4f$-shells of the magnetic atoms within a nanoscopic conductor such as a molecule or nanocontact which 
is coupled to semi-infinite electrodes while the rest of the system is treated on a static 
mean-field level by the local density approximation (LDA). This approach, can also be viewed as an extension of the early work 
of one of us\cite{Jacob:thesis:07} which ignored dynamical correlations of open atomic shells. 
In the limiting case where the correlated region reduces to a single atom, our approach reduces 
to our previous work which treated a single magnetic impurity in a metallic nanocontact\cite{Jacob:prl:09}. 
When the device region contains several correlated atoms in close proximity, the Molecular DMFT 
treatment is essential since the effective bath of each correlated atoms obeys a self-consistency 
condition which involves the whole device region.

Our work is similar in spirit to the recently presented dynamical vertex approximation 
for nanoscopic systems (nano-D$\Gamma$A) by A. Valli {\it et al}.\cite{Valli:prl:10} 
This approach can in principle treat non-local self-energy effects. In a related work, 
Florens introduced a nano-DMFT approach whereby a correlated system is approximated by 
embedding it in a model geometry having a tree like structure.\cite{Florens:prl:07}
However, both methods\cite{Valli:prl:10,Florens:prl:07} have been implemented in the 
context of the Hubbard model while we demonstrate that the existing Molecular DMFT 
technology can treat a realistic nanocontact.

This paper is organized as follows. In Sec. \ref{sec:method} we describe the Molecular DMFT method 
for nanoscopic conductors. In Sec. \ref{sec:results} we apply the method to small Ni nanocontacts
connected to Cu nanowires, and discuss the results. Finally, we conclude this paper with a 
general discussion of the method and of the results in Sec. \ref{sec:conclusions}. 

\section{Method}
\label{sec:method}

\begin{figure}
  \begin{center}
    \includegraphics[width=\linewidth]{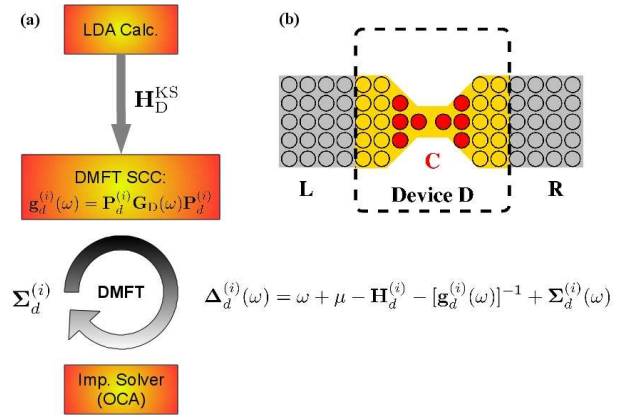}
  \end{center}
  \caption{(Color online)
    (a) Molecular DMFT self-consistency cycle applied to nanoscopic conductors.
    (b) Division of the system into left (L) and right (R) electrodes, and 
    the central device region (D) that hosts the strongly correlated subspace 
    (C) consisting of the strongly correlated $d$-orbitals of the magnetic atoms 
    shown as red (dark grey) circles.
  }
  \label{fig:method}
\end{figure}

We consider a nanoscopic conductor 
bridging two semi-infinite metal wires
as shown schematically in Fig. \ref{fig:method}(b).
As indicated, the nanoscopic conductor contains magnetic 
atoms that give rise to dynamic electron correlations 
due to the strongly interacting $3d$-electrons. 
The nanoscopic conductor could be for example a molecule,
a nanowire, a nano-cluster, or simply an atomic-size 
constriction in the metal wire.

In order to describe the dynamic correlations that arise from the 
$3d$-shells of the magnetic atoms we adapt the LDA+DMFT method to 
the case of nanoscopic conductors.
To this end it is convenient to divide the system into three parts 
as shown in the upper right panel of Fig. \ref{fig:method}: Two 
semi-infinite metallic leads L and R, and the central device region 
(D) which contains the nanoscopic conductor and the magnetic atoms with 
the strongly interacting $3d$-shells, as well as a sufficient 
part of the two metallic leads so that the electronic structure of 
the leads has relaxed to that of bulk (i.e. infinite) electrodes. 
The correlated subspace (C) that will be described on the level 
of DMFT is thus given by the direct sum of the $3d$-subspaces
of all magnetic atoms (M) in the device region.

The effective one-body Hamiltonians of the device region and leads are 
obtained from DFT calculations on the level of LDA. 
Here we use the supercell approach (see App. \ref{app:supercell}) to 
obtain the effective Kohn-Sham (KS) Hamiltonians of each part of the 
system prior to the dynamical treatment of the impurity $d$-shell and 
the transport calculations. The electronic structure of the device region 
is calculated with the {\sc Crystal06} {\it ab initio} electronic structure program 
for periodic systems \cite{Crystal:06} by defining a one-dimensional periodic 
system consisting of the device region as the unit cell. 

The device Hamiltonian $\mathbf{H}_{\rm D}$ is then obtained from the converged KS 
Hamiltonian of the unit cell of the periodic system. In the same way, the unit 
cell Hamiltonians $\mathbf{H}^0_{\rm L/R}$ and hoppings $\mathbf{V}_{\rm L/R}$
between unit cells of the left and right leads can be extracted from calculations 
of infinite nanowires with finite width since the electronic structure in the 
semi-infinite leads has relaxed to that of an infinite nanowire. 

The strong electron correlations of the $3d$-shells of the magnetic atoms are 
captured by adding a Hubbard-type interaction term to the one-body Hamiltonian 
within the correlated subspace $d$ of each of the magnetic atoms $i$:
\begin{equation}
  \label{eq:H_U}
  \hat{\mathcal{H}}_{U}^{(i)} = \frac{1}{2} \sum_{\stackrel{\alpha_1,\beta_1,\alpha_2,\beta_2}{\sigma_1,\sigma_2}} 
  U^{(i)}_{\alpha_1\alpha_2\beta_1\beta_2} \,
  \hat{d}_{i\alpha_1\sigma_1}^\dagger\hat{d}_{i\alpha_2\sigma_2}^\dagger 
  \hat{d}_{i\beta_2\sigma_2}\hat{d}_{i\beta_1\sigma_1}
\end{equation}
where $\hat{d}^\dagger_{i\alpha\sigma}$($\hat{d}_{i\alpha\sigma}$) creates (annihilates) 
an electron with spin $\sigma$ in the $3d$-orbital $\alpha$ on atom $i$. 
$U_{\alpha_1\alpha_2\beta_1\beta_2}^{(i)}$ are the matrix elements of the effective 
(i.e. screened) Coulomb interaction of the $3d$-electrons which is smaller than 
the bare Coulomb interaction due to the screening by the conduction electrons. 
Here we take a model interaction taking into account only the direct Coulomb
repulsion $U$ between electrons (i.e. $U_{\alpha\beta\alpha\beta}\equiv U$) 
and the Hund's rule coupling $J$ (i.e. $U_{\alpha\beta\beta\alpha}\equiv J$ for $\alpha\ne\beta$). 
For $3d$ transition-metal elements in bulk materials the repulsion $U$ is typically around 
2-3~eV and $J$ is around 0.9~eV.\cite{Grechnev:prb:07} 
However, due to the lower coordination of the atoms in the contact region or 
molecule the screening of the direct interaction should be somewhat reduced 
compared to bulk. Here we take $U=5$~eV and $J=0.9$~eV as in our previous work.\cite{Jacob:prl:09}

The Coulomb interaction within the correlated $3d$ subspaces of the magnetic atoms 
has already been taken into account on a static mean-field level in the effective 
KS Hamiltonian of the device. Therefore the KS Hamiltonian within each $3d$ subspace
has to be corrected by a double-counting correction term, i.e.
$\mathbf{H}_{d}^{(i)}\equiv \mathbf{P}_d^{(i)}\mathbf{H}_{\rm D}^{\rm KS}\mathbf{P}_d^{(i)}-\mathbf{H}_{\rm dc}^{(i)}$
where $\mathbf{P}_d^{(i)}$ is the projection onto the $3d$-subspace of atom $i$.\cite{foot2}
Here we use the standard expression \cite{Petukhov:prb:03}
\begin{equation}
  \mathbf{H}_{\rm dc}^{(i)} = [U(N_d^{(i)}-\frac{1}{2})-\frac{1}{2}J(N_d^{(i)}-1)]\times\mathbf{P}^{(i)}_{d}
\end{equation}
where $N_d^{(i)}$ is the occupation of the $3d$-shell of atom $i$.

The central quantity is the Green's function of the device region:
\begin{equation}
  \label{eq:GD}
  \mathbf{G}_{\rm D} = \left( (\omega+\mu)\mathbf{S}_{\rm D}
  -\mathbf{H}_{\rm D}
    +\mathbf{H}_{\rm dc}
    -\mathbf\Sigma_{\rm C}-\mathbf\Sigma_{\rm L}
    -\mathbf\Sigma_{\rm R}\right)^{-1}
\end{equation}
where $\mathbf\Sigma_{\rm C}$ is the electronic self-energy that describes 
the dynamic electron correlations of the electrons within the correlated 
subspace C, and $\mathbf{H}_{\rm dc}$ is the double-counting correction within
the entire correlated subspace C, i.e., $\mathbf{H}_{\rm dc}\equiv\sum_{i\in M}\mathbf{H}_{\rm dc}^{(i)}$.
$\mathbf{S}_{\rm D}$ is the overlap matrix taking into account the non-orthogonality of basis set.
\cite{foot3}
$\mu$ is the chemical potential, $\mathbf\Sigma_{\rm L}$ and $\mathbf\Sigma_{\rm R}$ 
are the so-called lead self-energies\cite{foot1} which describe the coupling of the device to 
the semi-infinite leads L and R, respectively. These can be calculated from the 
effective one-body Hamiltonians of the leads by iteratively solving the Dyson equation
for the lead self-energies, eq. (\ref{eq:Dyson}).


The central assumption of DMFT is that the inter-site correlations, i.e.
the correlations between electrons located on different atoms can be neglected.
In that case the electron correlation self-energy $\mathbf\Sigma_{\rm C}$ becomes 
block-diagonal, and each block corresponds to the self-energy $\mathbf\Sigma_{d}^{(i)}$
of the correlated $3d$ subspace of a magnetic atom $i$: 
$\mathbf\Sigma_{d}^{(i)}=\mathbf P_{d}^{(i)} \mathbf\Sigma_{\rm C} \mathbf P_{d}^{(i)}$.
The self-energies $\mathbf\Sigma_{d}^{(i)}$ and hence the overall self-energy $\mathbf\Sigma_{\rm C}$
can now be determined by mapping onto a generalized Anderson impurity problem
for each correlated subspace, described by the following Green's function:
\begin{equation}
  \label{eq:gd}
  \mathbf{g}_{d}^{(i)}(\omega) = \left(\omega+\mu-\mathbf{H}_{d}^{(i)}
    -\mathbf\Sigma_{d}^{(i)}(\omega)-\mathbf\Delta_{d}^{(i)}(\omega)\right)^{-1}
\end{equation}
where we have introduced the so-called hybridization function $\mathbf\Delta_{d}^{(i)}$ 
which describes the coupling of the correlated subspace with the rest of the
system. The hybridization function is determined by the DMFT self-consistency 
condition (DMFT-SCC):
\begin{equation}
  \label{eq:DMFT-SCC}
  \mathbf{g}_{d}^{(i)}(\omega) = \mathbf{P}_{d}^{(i)} \mathbf{G}_{\rm D}(\omega) \mathbf{P}_{d}^{(i)}
\end{equation}
It follows that the hybridization function is given by:
\begin{equation}
  \label{eq:Delta}
  \mathbf\Delta_{d}^{(i)}(\omega) = \omega+\mu-\mathbf{H}_{d}^{(i)}-[\mathbf{g}_{d}^{(i)}(\omega)]^{-1}-\mathbf\Sigma_{d}^{(i)}(\omega)
\end{equation}

The equations (\ref{eq:GD}-\ref{eq:Delta}) define the Molecular DMFT self-consistency cycle for the
calculation of the self-energies $\mathbf\Sigma_{d}^{(i)}$: One starts with 
the effective one-body Hamiltonian $\mathbf{H}_{\rm D}$ obtained from the LDA calculation and an initial 
guess for the local correlation self-energies $\mathbf\Sigma_{d}^{(i)}$ (usually zero). This 
allows one to calculate the device Green's function, eq. (\ref{eq:GD}), and consequently the
projection $\mathbf{g}_{d}^{(i)}$. Thus one obtains the Hybridization functions $\mathbf\Delta_{d}^{(i)}(\omega)$
which together with the on-site energy levels $\mathbf{H}_{d}^{(i)}$ and the Coulomb interaction $U$, $J$
defines the Anderson impurity problem which can be solved by an impurity solver, and one
obtains a new self-energy $\mathbf{\Sigma}_{d}^{(i)}$.
The DMFT self-consistency cycle is illustrated in Fig. \ref{fig:method}(a).

Solving the generalized Anderson impurity problem is a difficult task, and at present there is
no universal impurity solver that works efficiently and accurately in all parameter regimes.
Here we make use of impurity solvers based on an expansion in the hybridization strength given 
by $\Delta_{d}^{(i)}(\omega)$ around the atomic limit. The starting point is an exact diagonalization
of the (isolated) impurity subspace i.e., the $3d$-shell of the magnetic atom given by the
interacting Hamiltonian $\hat{\mathcal H}_{d}^{(i)}+\hat{\mathcal H}_{U}^{(i)}$. 
The hybridization of the impurity subspace with the rest of the system (given by the hybridization 
function $\Delta_{d}^{(i)}(\omega)$) is then treated perturbatively. 

The so-called Non-Crossing Approximation\cite{Bickers:rmp:87} (NCA) is a self-consistent perturbation expansion to lowest 
order in the hybridization strength. NCA only takes into account bubble diagrams describing 
hopping processes where an electron or hole hops into the impurity at some time and then out 
at a later time (see Fig. \ref{fig:nca+oca} in App. \ref{app:nca+oca}). 
The OCA\cite{Pruschke:zphysb:89,Haule:prb:01} improves on the NCA by taking into account 
second order diagrams where two additional electrons (holes) are accommodated on the impurity 
at the same time as shown in Fig. \ref{fig:nca+oca}. 
OCA improves considerably many of the shortcomings of NCA\cite{Bickers:rmp:87}: It substantially improves the width
of the Kondo peak and hence the Kondo temperature which now are only slightly underestimated.
It also corrects the asymmetry of the Kondo peak. For very low temperatures ($T \ll T_K$), however, 
the height of the Kondo peak is still overestimated, and the Fermi liquid behavior at zero 
temperature is not recovered.\cite{Haule:prb:01} 

Hence, OCA is a reasonable approximation for solving the generalized impurity problem
as long as the temperatures are not too low (i.e. more than one order of magnitude below $T_K$).
In App. \ref{app:nca+oca} we give a brief introduction to the NCA and OCA impurity solvers.
A detailed description of the NCA and OCA methods can be found e.g. in Refs.
\onlinecite{Kotliar:rmp:06,Haule:prb:10,Hewson:book,Haule:prb:01,Pruschke:zphysb:89,Bickers:rmp:87}.


The current through a strongly interacting region can be calculated 
exactly by the Meir-Wingreen formula \cite{Meir:prl:92}. But in order to 
apply the Meir-Wingreen result one has to solve the impurity problem
out of equilibrium which is a difficult task that has only been accomplished 
very recently and only in the context of model Hamiltonians\cite{Werner:prb:10,Dirks:pre:10}.
However, Meir and Wingreen also showed that for low temperatures and small bias 
voltages the Meir-Wingreen expression is well approximated by the much simpler 
Landauer formula:\cite{Landauer:philmag:70}
\begin{equation}
  I(V) = \frac{2e}{h}\times\int_{0}^{eV}d\omega \, T(\omega)
\end{equation}
where $T(\omega)$ is the Landauer transmission function and where we have 
assumed an asymmetric voltage drop $V$ about the device region.
Thus the conductance is simply given by the Landauer transmission function:
\begin{equation}
  \mathcal{G}(V)=\frac{\partial I}{\partial V}(V) = \frac{2e^2}{h}\times T(eV)
\end{equation}
The latter can be calculated from the (equilibrium) device Green's function:
\begin{equation}
  T(\omega) = {\rm Tr}[ \mathbf\Gamma_{\rm L}(\omega) \mathbf{G}_{\rm D}^\dagger(\omega) 
    \mathbf\Gamma_{\rm R}(\omega) \mathbf{G}_{\rm D}(\omega) ]
\end{equation}
where $\mathbf\Gamma_{\rm L/R}$ are the so-called coupling matrices which describe
the coupling to the leads, and can be calculated from the lead self-energies by
$\mathbf\Gamma_{\rm L/R} = i ( \mathbf\Sigma_{\rm L/R} - \mathbf\Sigma^\dagger_{\rm L/R} )$.

\section{Results and Discussion}
\label{sec:results}

\begin{figure}
  \begin{tabular*}{\linewidth}{c@{\extracolsep\fill}c}
     (a) & (b) \\
     \includegraphics[width=0.45\linewidth]{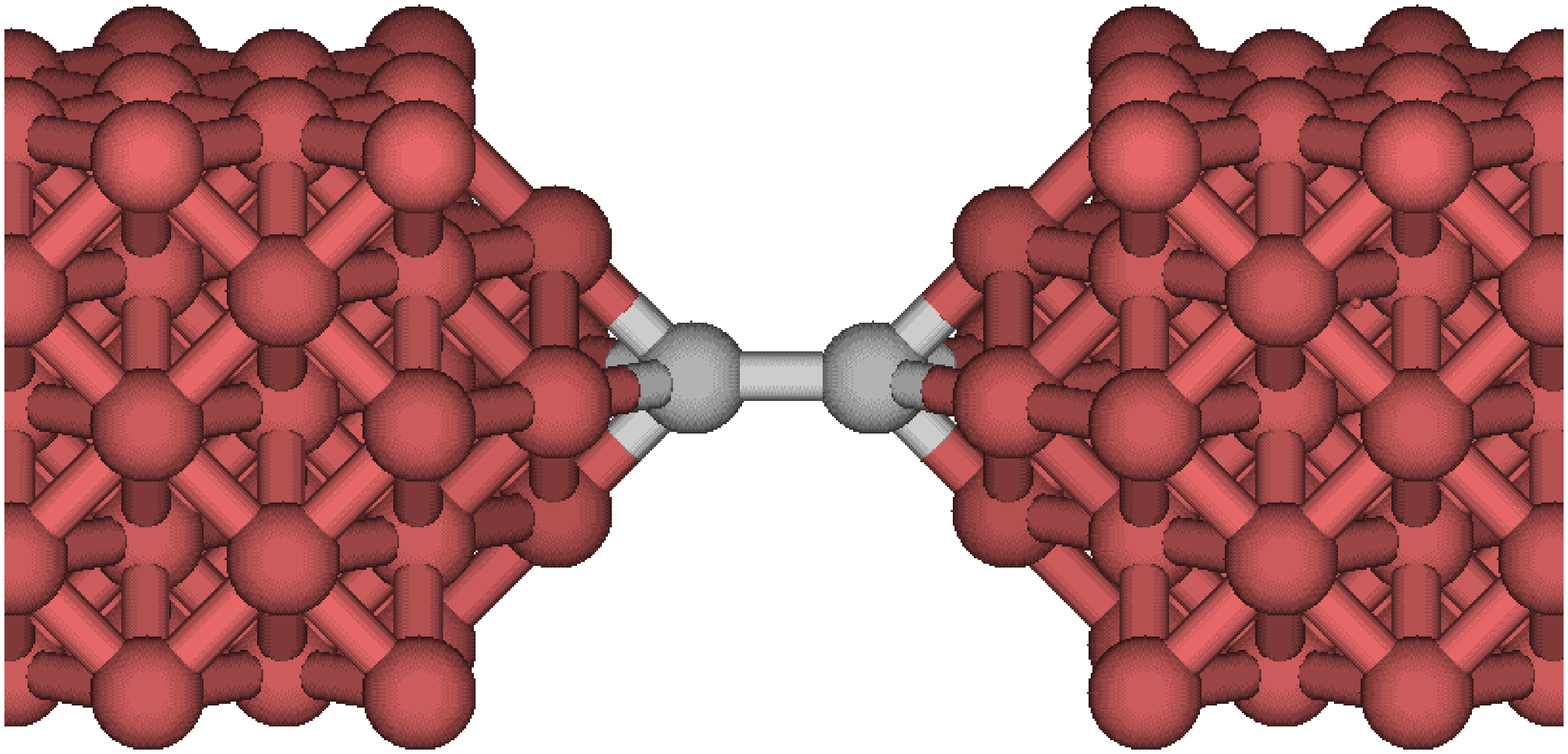} &
     \includegraphics[width=0.45\linewidth]{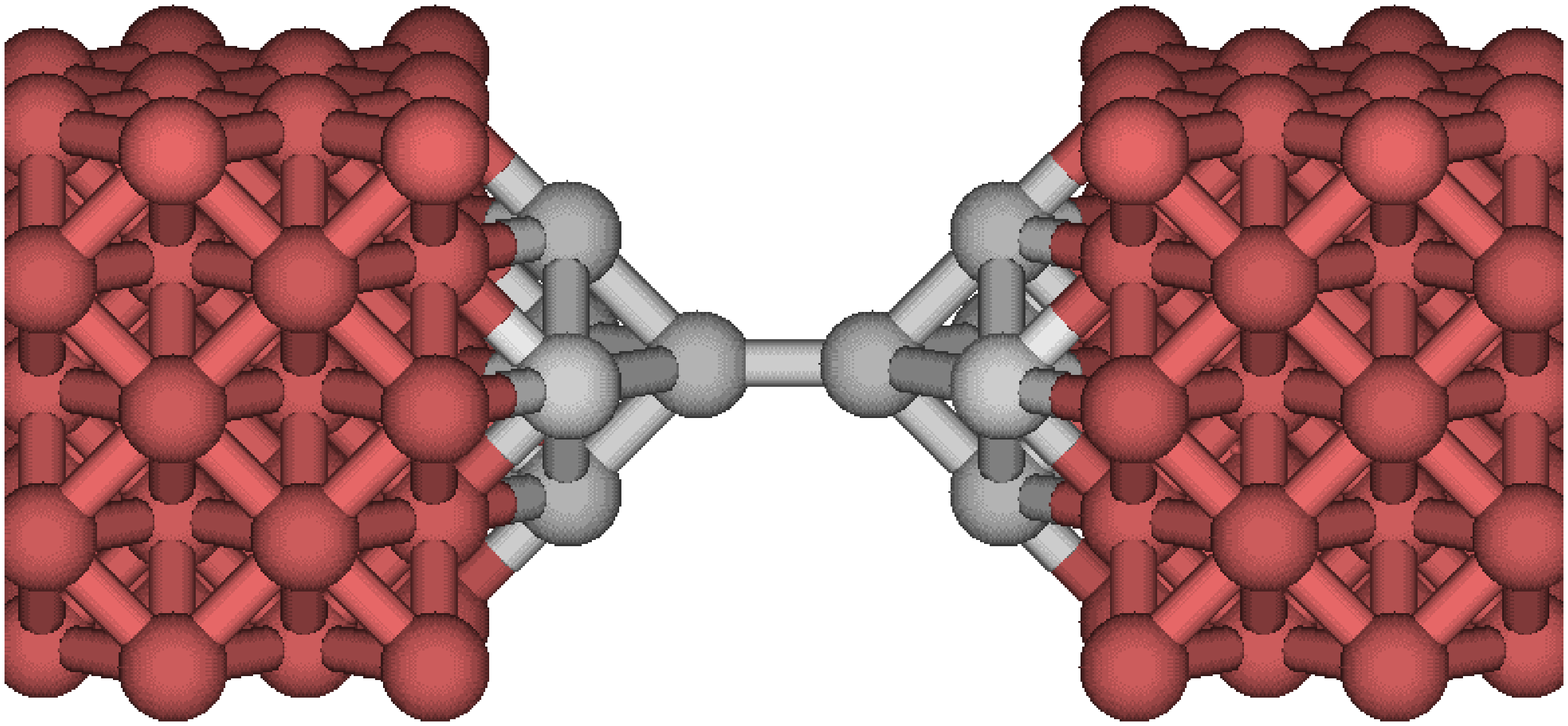}
   \end{tabular*}
  \caption{(Color online) (a) Ni dimer between the tips of 
    two Cu nanowires grown in the 001 direction of bulk Cu. 
    (b) Ni nanocontact consisting of 10 atoms between two 
    Cu 001 nanowires. Ni atoms are shown in light grey 
    (light grey) and Cu atoms are shown in red (dark grey).
  }
  \label{fig:nanocontacts}
\end{figure}

In the following we demonstrate the above developed
Molecular DMFT method for the two idealized Ni nanocontacts
with Cu nanowires as electrodes shown in Fig. \ref{fig:nanocontacts}.
Ni nanocontacts have been a subject of intense research in the
last decade as prospective ingredients for nanoscale spintronics
devices.\cite{Viret:prb:02,Untiedt:prb:04,Keane:apl:06,Bolotin:nl:06,
Jacob:prb:05,Smogunov:prb:06,Jacob:prb:06b,Jacob:prb:08}
Here we consider the paramagnetic phase, i.e. the 
self-energies and the Green's functions are spin-degenerate.
Breaking of the spin-symmetry by an external magnetic field
or by spin-polarized electrodes is not taken into account.

\subsection{Ni dimer between Cu nanowires}
\label{sec:dimer}

First, we consider a dimer of Ni atoms between two semi-infinite Cu 
nanowire electrodes as shown in Fig. \ref{fig:nanocontacts}a. The 
nanowires are grown in the 001 direction of bulk Cu. The distance 
between the two Ni tip atoms is 2.4~\r{A}, all other distances are
those of bulk Cu. For the sake of simplicity, we have not relaxed 
the atomic positions. Due to the highly idealized geometry both Ni 
atoms are equivalent. Hence in each step of the Molecular DMFT 
self-consistency we only have to solve the impurity problem once. 

\begin{figure}
  \includegraphics[width=\linewidth]{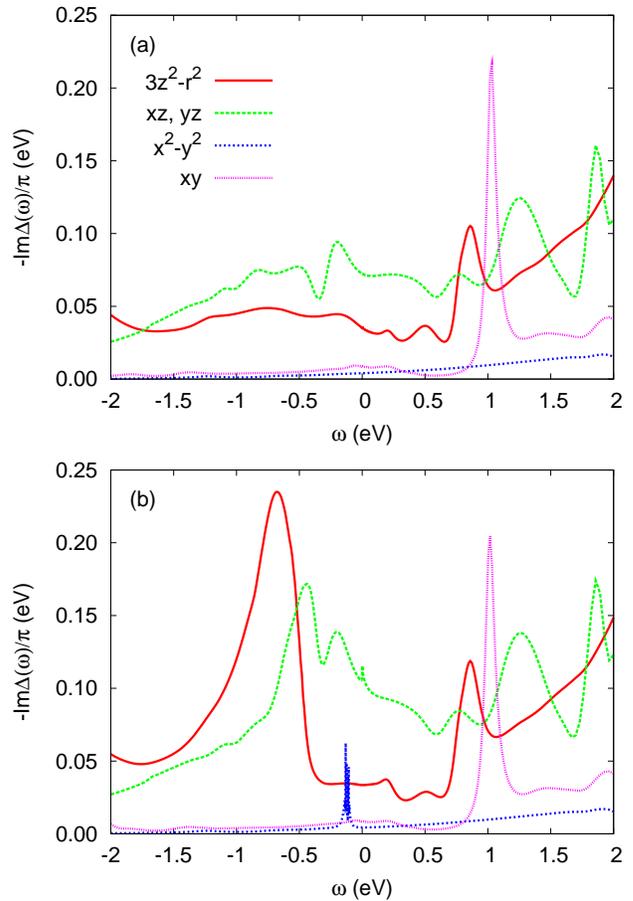}
  \caption{
    Imaginary part of hybridization functions for different 
    Ni $3d$-orbitals in the dimer geometry shown in Fig. 
    \ref{fig:nanocontacts}(a) in the first step (a) and in the 
    last step (b) of the self-consistent Molecular DMFT calculation
    at low temperature ($T=12$~K).
  }
  \label{fig:dimer:imdelta}
\end{figure}

In spite of the highly symmetric situation the Ni $3d$-orbitals 
split into four different symmetry groups. As can be seen from 
Fig. \ref{fig:dimer:imdelta}(a) which shows the hybridization 
function in the first step of the self-consistent DMFT procedure
(where the self-energy $\Sigma_d(\omega)$ is zero), the hybridization 
functions are quite different for each of the four symmetry groups. 
This situation is different from that of the corresponding bulk systems 
where the hybridization functions for each correlated orbital are usually 
very similar due to the highly isotropic closed packed crystal structures.
In the geometry considered here, the doubly-degenerate $3d_{xz}$- and $3d_{yz}$-
orbitals have the strongest hybridization around the Fermi-level with the 
rest of the system. Also the $3d_{3z^2-r^2}$-orbital has an appreciable hybridization.
The hybridization of the $3d_{xy}$- and $3d_{x^2-y^2}$-orbitals on the other
hand are smaller by at least one order of magnitude.

Figure \ref{fig:dimer:imdelta}(b) shows the hybridization function after
self-consistency has been reached in the DMFT calculation. Now the 
converged self-energy $\Sigma_d(\omega)$ is non-zero in general. 
We can see that the DMFT self-consistency has a considerable effect 
on the hybridization function for most of the Ni $3d$-orbitals. 
For example, the hybridization of the degenerate $3d_{xz}$- and 
$3d_{yz}$-orbitals around the Fermi level is strongly increased. 
Moreover a sharp peak appears right at the Fermi level. Additional
features also arise in the hybridization functions of the other
$3d$-orbitals. 
The additional features in the $3d$-hybridization functions of one 
Ni atom stem from the corresponding features (induced by the 
on-site interactions) in the $3d$-spectral density of the other 
Ni atom.

\begin{figure}
  \includegraphics[width=\linewidth]{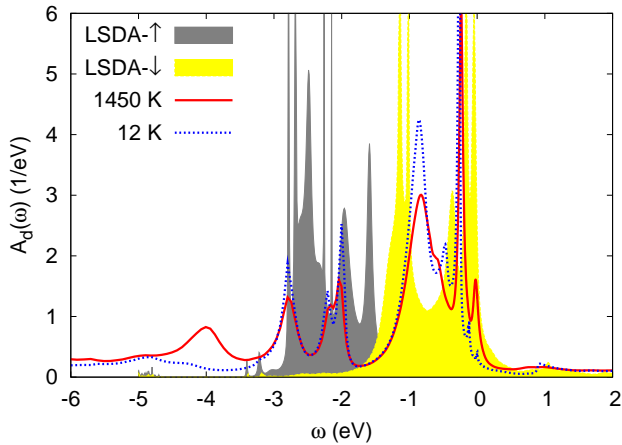}
  \caption{
    Ni $3d$-spectral function for the Ni dimer between 
    Cu nanowires shown in Fig. \ref{fig:nanocontacts}(a) 
    calculated with Molecular DMFT at two different temperatures
    on the one hand and calculated with LSDA on the other 
    hand.
    \label{fig:dimer:spectra-temp}
  }
\end{figure}

\begin{figure}
  \includegraphics[width=\linewidth]{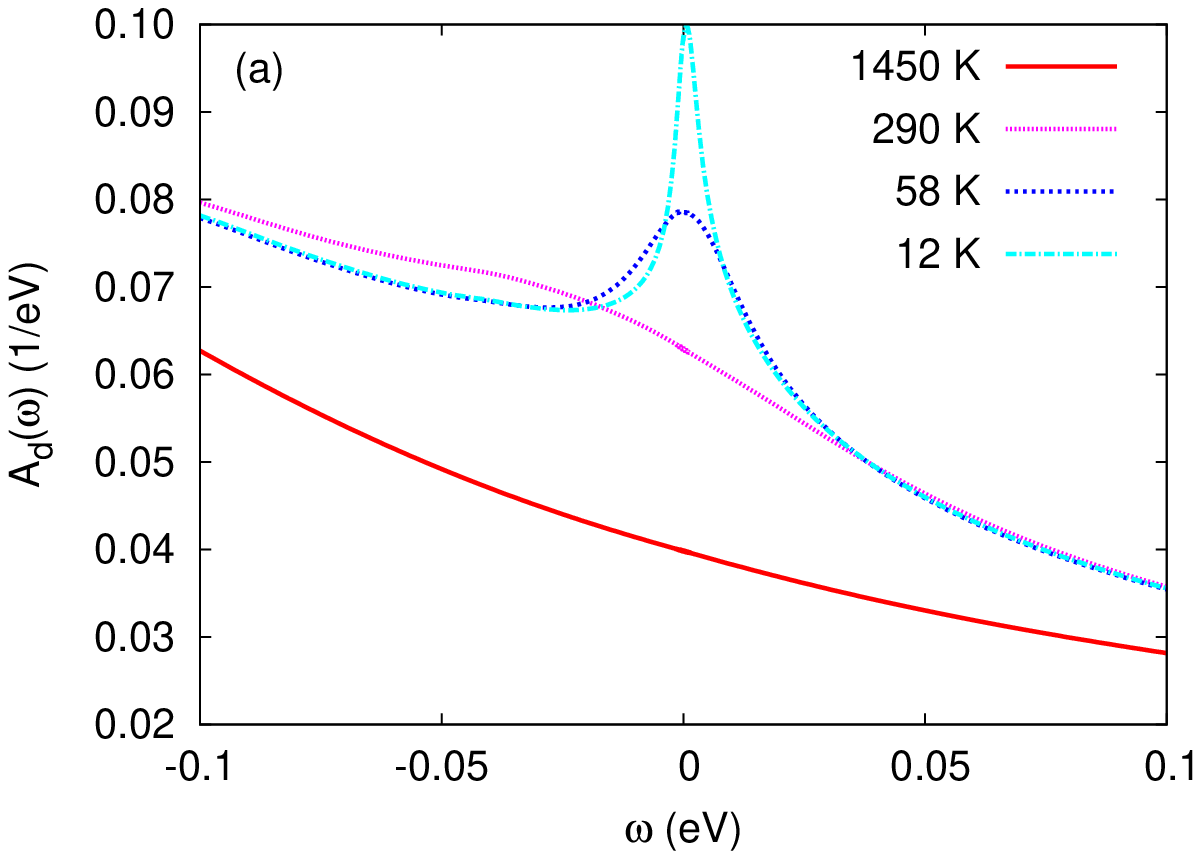}
  \includegraphics[width=\linewidth]{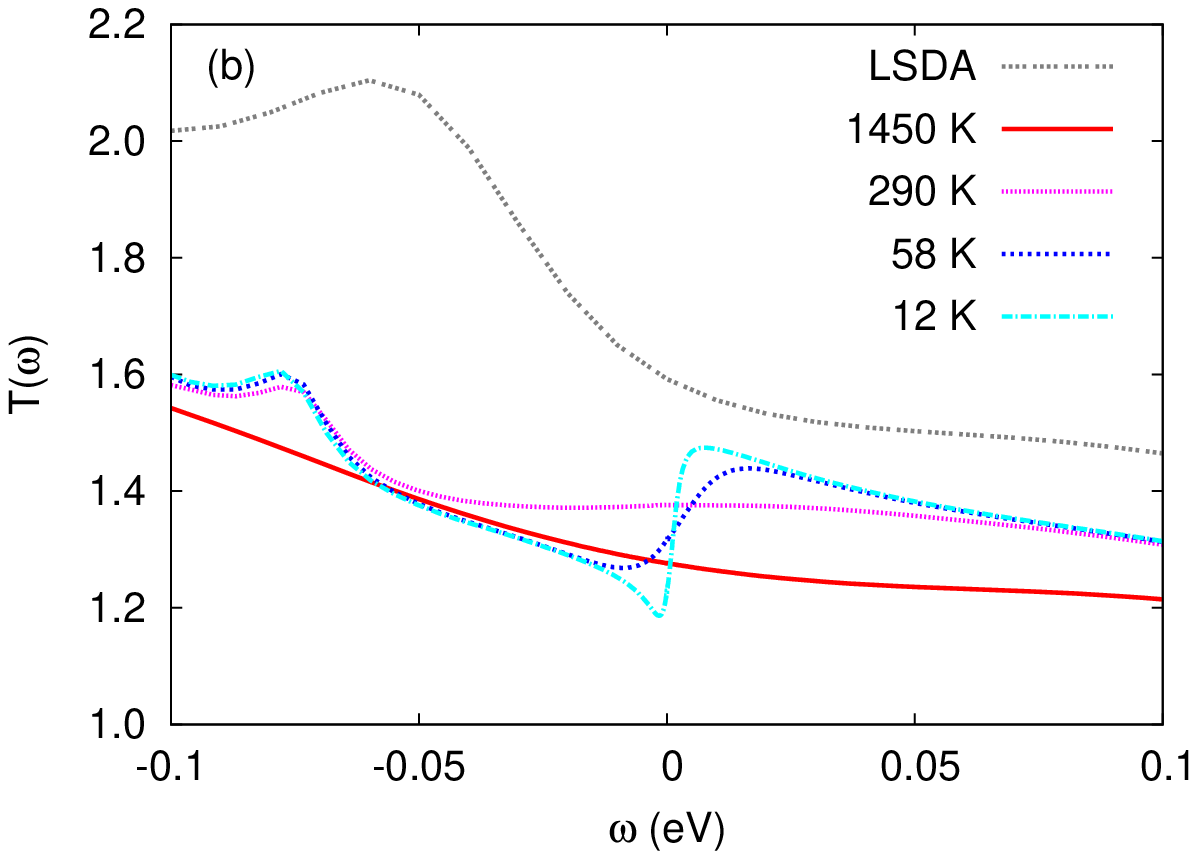}
  \caption{
    (a) Spectral function of Ni $3d_{xz}$- and $3d_{yz}$-orbitals 
    near the Fermi level for different temperatures. (b) Correlated 
    transmission function calculated with Molecular DMFT at different 
    temperatures compared to LSDA transmission function near the Fermi 
    level.
  }
  \label{fig:dimer:kondo-fano}
\end{figure}

In Fig. \ref{fig:dimer:spectra-temp} we show the temperature dependence 
of the Ni $3d$-spectrum calculated with Molecular DMFT and compare them 
to the spectrum calculated with LSDA. The DMFT spectra are somewhat smother
than the LSDA spectra due to the finite lifetime broadening of the single-particle
states by the electron-electron interactions.
Most importantly, the Molecular DMFT spectrum shows a strong temperature dependence 
which cannot be captured by a static mean-field treatment like LSDA.
For example, at low temperatures a small peak forms right at the Fermi 
level in the Molecular DMFT spectrum. This is a quasi-particle peak that originates 
from the two degenerate $3d_{xz}$- and $3d_{yz}$-orbitals as can be seen from 
Fig. \ref{fig:dimer:kondo-fano}(a). Strictly speaking, it is not a Kondo peak 
since the occupation of the two orbitals is around 3.65, and hence these orbitals
are in the so-called mixed-valence regime where in addition to the spin-fluctuations
also charge fluctuations take place (see e.g. the book by Hewson \onlinecite{Hewson:book}
for a detailed discussion). However, we would like to emphasize here that the 
corresponding magnetic moment is nevertheless screened by the conduction 
electrons. 

We can find the weight $Z$ and the linewidth $\Gamma$ of the quasi-particle
by fitting the peak to a Lorentzian. Fig. \ref{fig:dimer:kondo-fano}(a) suggests 
that the $3d_{xz}$-,$3d_{yz}$-spectral function near the Fermi level can be fitted 
to a weighted Lorentzian plus a linear function:
\begin{equation}
  \label{eq:lorentzian}
  A_d(\omega) = a\omega+b + \frac{Z}{\pi}\frac{\Gamma/2}{\omega^2 + \Gamma^2/4}
\end{equation}
We find a very tiny quasi-particle weight $Z$ of less than 0.1\%, and a width $\Gamma$ 
corresponding to a temperature of about 130~K. This is the critical temperature below 
which the quasi-particle can be observed (and which in the case of Kondo effect is 
called Kondo temperature). 
Note that the width of the quasi-particle is somewhat enhanced by the DMFT
self-consistency compared to the case without DMFT (not shown).

The quasi-particle peak leads to a corresponding Fano feature in the 
transmission function as can be seen in Fig. \ref{fig:dimer:kondo-fano}(b)
which shows the transmission function for small energies around the Fermi level.
Hence our calculations show that the low bias conductance of the Ni dimer
between Cu electrodes features a Fano-lineshape due to a quasi-particle peak
in the Ni $3d$-spectral function at low temperatures.

Table \ref{tab:occ} lists the individual occupations of all Ni $3d$-orbitals 
and the corresponding effective energy levels that are obtained from the KS
energy levels plus the real part of the hybridization function at zero frequency.
We can see that all orbitals apart from the $d_{xy}$-orbital have mixed valences
and are almost full. The $d_{xy}$-orbital on the other hand is basically half-filled.
hence this orbital would be a candidate for a true Kondo effect, i.e. screening
of the magnetic moment by spin-fluctuations only. However, the hybridization of
this orbital is very low, as can be seen from Fig. \ref{fig:dimer:imdelta}. Hence the
Kondo temperature for this orbital is very low so that the orbital does not enter
the Kondo regime at the temperatures considered here but is in the local moment regime.

\subsection{Ni nanocontact between Cu electrodes}

\begin{figure}
  \includegraphics[width=\linewidth]{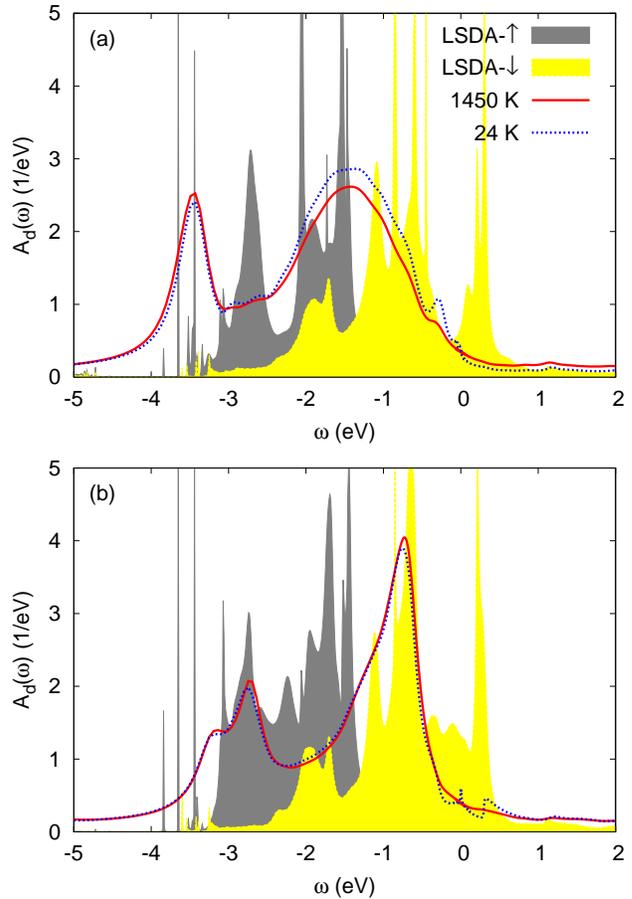}
  \caption{
    (a) $3d$-spectral function of a Ni atom 
    at the base of one of the pyramids of the
    Ni nanocontact between Cu nanowires
    shown in Fig. \ref{fig:nanocontacts}(b)
    calculated on the one hand with Molecular DMFT
    at two different temperatures and with
    LSDA on the other hand.
    (b) Same as (a) but for the tip atoms
    of the Ni nanocontact.
    \label{fig:nanoc:spectra-temp}
  }
\end{figure}

Now we turn to the slightly more complicated case
of the Ni nanocontact consisting of 10 atoms between
two Cu nanowires as shown in Fig. \ref{fig:nanocontacts}(b).
As before the nanowires are grown in the 001 direction of 
bulk Cu. The distance between the two Ni tip atoms is 
2.4~\r{A} while all other distances are those of bulk Cu. 
As before the two Ni tip atoms are both equivalent. On the 
other hand, the eight outer atoms of the Ni nanocontact are 
not equivalent with the tip atoms, but are equivalent among
themselves. Hence we have to solve two different impurity 
problems in each step of the Molecular DMFT self-consistency 
cycle.

\begin{figure}
  \includegraphics[width=\linewidth]{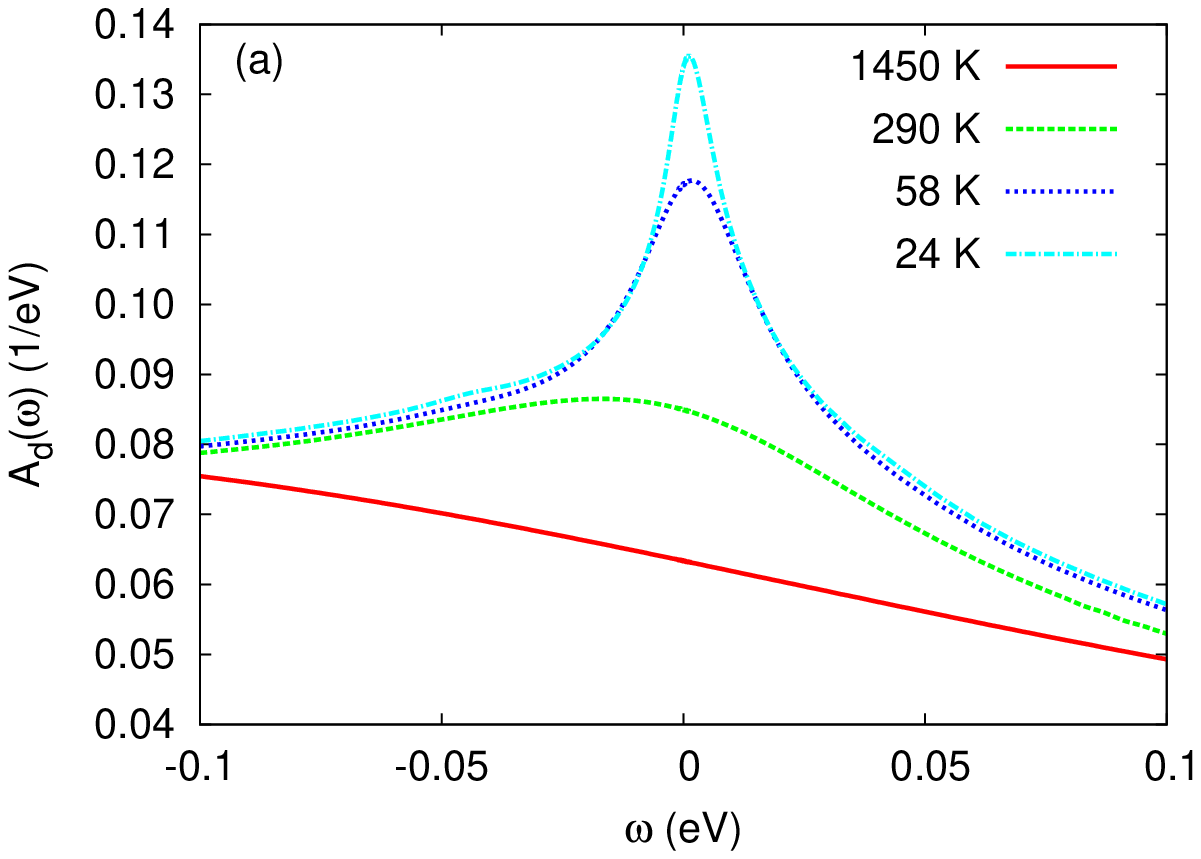}
  \includegraphics[width=\linewidth]{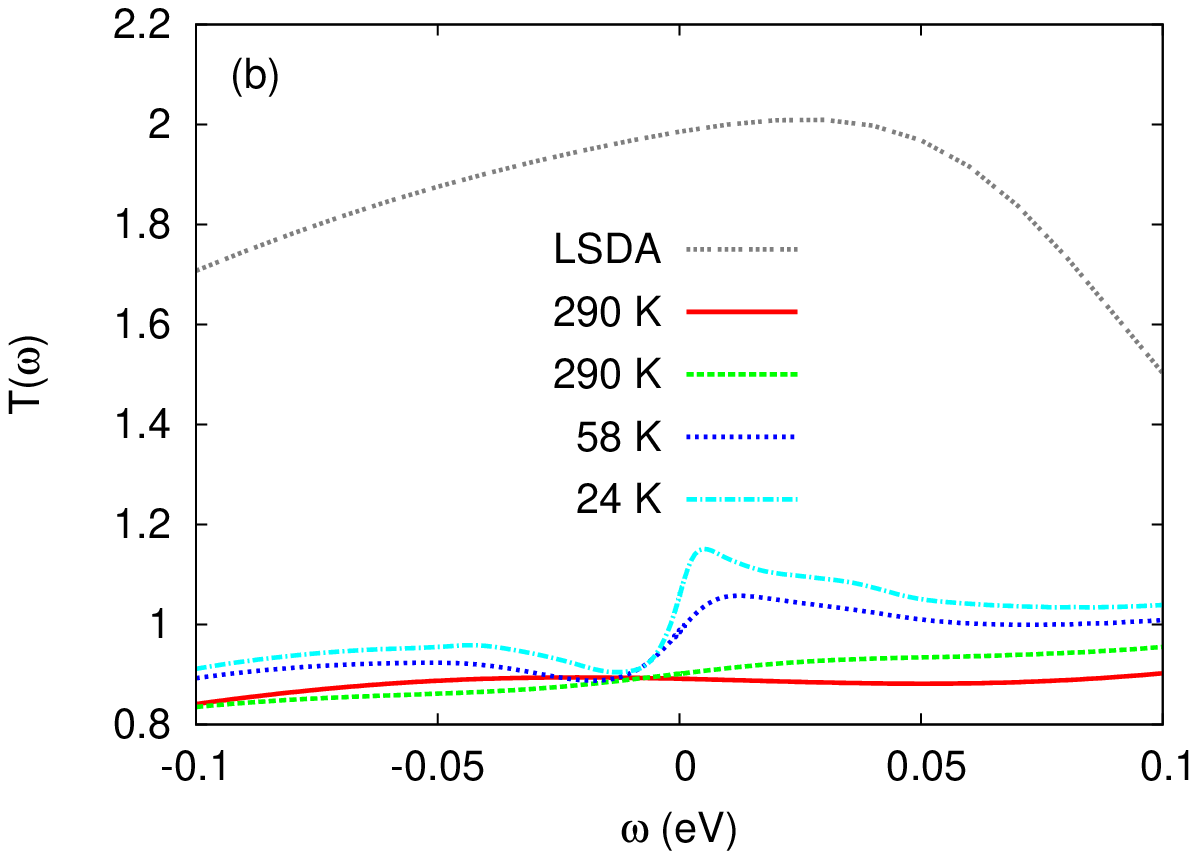}
  \caption{
    (a) Spectral function of $3d_{xz}$- and $3d_{yz}$-
    orbitals of Ni tip atoms near the Fermi level for 
    different temperatures.
    (b) Correlated transmission function at different temperatures 
    compared to LSDA transmission function for energies around 
    the Fermi level.
  }
  \label{fig:nanoc:kondo-fano}
\end{figure}

In Fig. \ref{fig:nanoc:spectra-temp}, we show the $3d$-spectral
function of (a) the base atoms and (b) the tip atoms of the Ni
nanocontact calculated on the one hand with Molecular DMFT at 
different temperatures, and on the other hand with DFT on the 
level of the LDA. We see that in both cases the Molecular DMFT
spectral densities are quite different from the LSDA ones. 
Both LSDA spectra feature a strong peak above the 
Fermi level for the minority electrons that is absent in
the Molecular DMFT spectra. 
Moreover, near the Fermi level, the Molecular DMFT spectrum of the tip atoms 
is strongly temperature dependent due to the formation of a quasi-particle 
peak at the Fermi level in the degenerate $3d_{xz}$- and $3d_{yz}$-levels 
for low temperatures. Fig. \ref{fig:nanoc:kondo-fano}(a) shows a close-up
of the formation of the quasi-particle peak in the spectral density of
the $3d_{xz}$- and $3d_{yz}$-levels. Fitting to a Lorentzian plus linear 
function, eq. (\ref{eq:lorentzian}), we now find a slightly increased 
quasi-particle weight $Z$ of 0.2\% and also a somewhat increased width 
$\Gamma$ corresponding to a critical temperature of 220~K as compared to
the case of the Ni dimer. 
The increased width $\Gamma$ and quasi-particle weight $Z$ is due to the 
increased imaginary part of the hybridization function (not shown) of the 
tip atoms near the Fermi level which stems from the $3d$-spectral density of the 
base atoms of the Ni pyramid. In the dimer case, the base atoms of the 
pyramids are Cu atoms where the $3d$-spectral density near the Fermi level
is negligible.

The occupation of the degenerate $3d_{xz}$- and $3d_{yz}$-orbitals is 3.73.
Hence these orbitals are in the mixed-valence regime rather than the Kondo 
regime as in the case of the Ni dimer. Therefore the quasi-particle peak is 
strictly speaking not a Kondo peak. However, as was said before, the magnetic 
moment of these orbitals is nevertheless screened in the mixed valence regime
by the spin- and charge-fluctuations.

Figure \ref{fig:nanoc:kondo-fano}(b) shows the transmission function near
the Fermi level calculated on the one hand with the Molecular DMFT method 
at different temperatures and on the other and with DFT on the level of the
LDA. As in the case of the dimer, the formation of the quasi-particle peak 
in the $3d_{xz}$- and $3d_{yz}$-orbitals at low temperatures leads to a 
temperature dependent Fano lineshape in the Molecular DMFT transmission 
function. The LSDA transmission function on the other hand does not show
such a behavior. On the contrary, it is rather flat and featureless at this
energy scale. Also note that the LSDA transmission is considerably higher
than the correlated transmission calculated with Molecular DMFT.
This is due to the correlations shifting a considerable part of the spectral 
weight of the $3d$-orbitals away from the Fermi level. Consequently, the 
overall contribution of the $3d$-orbitals to the transmission is higher in 
the case of the LSDA calculation than in the Molecular DMFT calculation.

The transmission as calculated with the Molecular DMFT approach, although
lower than the LSDA one, is still compatible with the broad peak roughly 
between 1 and 1.7$\times G_0$ in the experimentally measured conductance
histograms of Ni nanocontacts.\cite{Untiedt:prb:04} 

\begin{table}
  \begin{tabular*}{\linewidth}{c@{\extracolsep\fill}cccc}
    \hline
    \hline
    \rule{0ex}{2.5ex}
    &\multicolumn{2}{c}{Ni dimer} & \multicolumn{2}{c}{Ni$_{10}$-nanocontact}\\
    & $n_d$ & $\tilde\epsilon_d$ & $n_d$ & $\tilde\epsilon_d$ \\
    &       &  (eV)              &       &     (eV) \\ [0.5ex]
    \hline
    \rule{0ex}{2.5ex}
    $3z^2-r^2$ & $1.8801$ & $-5.2188$ & $1.9336$ & $-5.4886$ \\ [0.5ex] 
    $xz$, $yz$ & $3.6468$ & $-5.1296$ & $3.7395$ & $-5.3954$ \\ [0.5ex]
    $x^2-y^2$  & $1.9174$ & $-4.8121$ & $1.0096$ & $-5.1749$ \\ [0.5ex]
    $xy$       & $0.9943$ & $-4.7410$ & $1.9578$ & $-5.2706$ \\ [0.5ex]
    \hline
    \hline
  \end{tabular*}
  \caption{
    Orbital occupations $n_d$ (at low temperature) and effective energy levels 
    $\tilde\epsilon_d\equiv\epsilon_d+{\rm Re}\Delta_d(0)$ 
    of the $3d$-orbitals of the Ni tip atoms in the case of the Ni dimer geometry 
    and the Ni nanocontact consisting of 10 Ni atoms.
    $\epsilon_d$ denotes the KS energy levels before double-counting correction.
  }
  \label{tab:occ}
\end{table}

In Tab. \ref{tab:occ} the orbital occupations and effective energy-levels
of individual Ni $3d$-orbitals of a tip atom of the Ni nanocontact are shown.
The most striking difference with the Ni dimer (shown in the same table) is
that now the $d_{x^2-y^2}$-orbital is the highest energy orbital with half-filling
instead of the $d_{xy}$-orbital. This is a consequence of the different
environment of the tip atom in the 10-atom nanocontact geometry compared
to the dimer geometry. The presence of the $3d$-orbitals of the Ni atoms at 
the base of each pyramid near the Fermi levels change the hybridization functions
of the tip atoms accordingly. Note that here the DMFT self-consistency is essential 
since without the self-consistency the hybridization functions would be very similar 
to the dimer case.

\section{Summary and conclusions}
\label{sec:conclusions}

We have developed a method for calculating the
electronic structure and transport properties of nanoscopic conductors
that explicitly takes into account local dynamical correlations originating
from strong electron-electron interactions. Our method extends the established 
DFT based {\it ab initio} transport methodology for nanoscopic conductors
to include dynamic electron correlations originating from the 
strongly interacting $3d$-electrons of the transition metal atoms.
This is achieved by combining the DFT electronic structure calculations
of the nanoscopic conductor with a DMFT description of the strongly interacting 
$3d$-electrons in the device region. We thus obtain the correlated
Green's function of the nanoscopic conductor which allows to calculate
the electronic structure and the corresponding conductance in the low-bias
voltage regime.

We have demonstrated the method for two model systems, namely Ni nanocontacts
consisting of several atoms and connected to Cu leads. We find that the dynamic 
correlations of the strongly interacting Ni $3d$-electrons give rise to strongly 
temperature-dependent spectra due to the formation of a quasi-particle peak
at low temperatures. The quasi-particle peak gives rise to a temperature-dependent 
Fano-type lineshape in the low-bias conductance characteristics similar to those
measured in recent experiments with ferromagnetic nanocontacts.\cite{Calvo:nature:09}
Moreover, the critical temperatures of 120~K and 220~K for the formation of the 
quasi-particle peak is in quite good agreement with the broad distribution of 
Kondo temperatures around the average temperature of 250~K extracted from the 
Fano lineshapes in the low-bias conductance measurements of Ni nanocontacts
in that same experiment. 
Note that a quasi-particle peak at the Fermi level can in principle also be 
obtained in the GW approximation.\cite{Thygesen:prb:08} 
However, in order to capture the quasi-atomic features characteristic 
of the strong correlation regime such as Hubbard bands or satellites together 
with the concomitant renormalization of the quasi-particle, a non-perturbative
treamtent of the local part of the Coulomb interaction such as the Molecular DMFT
method presented here is necessary.

The quasi-particle peak obtained here, is strictly speaking, not a Kondo peak 
since the system is in the so-called mixed-valence regime where charge fluctuations 
take place in addition to the spin-fluctuations that lead to the Kondo effect. 
This hints at the possibility that the origin of the Fano lineshapes 
in the low-bias conductance of Ni nanocontacts measured experimentally need
not always be the Kondo effect. However, we would like to stress that also 
in the mixed-valence regime the magnetic moment of the corresponding orbitals 
would be screened. But to draw further conclusions in that matter, more 
realistic calculations are necessary taking into account the ferromagnetic
leads, and sampling over different contact geometries.

We have illustrated the Molecular DMFT method for the case of simple nanocontacts 
containing several transition metal atoms, but the Molecular DMFT approach is 
very general and can be applied to many systems of great theoretical and practical 
interest. For example, it can be used to treat large molecules in which one can isolate 
small clusters of correlated elements as for example in the fuel cell materials of Tard 
{\it et. al.}\cite{Tard:nature:05} 

The Molecular DMFT method allows to explicitly incorporate strong dynamical correlations 
within the established DFT based transport methodology for nanoscopic conductors.
Our calculations show that dynamical correlations originating from the strongly 
interacting shells of magnetic atoms can alter the electronic structure and transport 
properties of nanoscopic conductors significantly.

\section*{Acknowledgments}

We are grateful for fruitful discussions with Erio Tosatti and Karsten Held.
DJ acknowledges funding by DAAD during his stay at the Department of Physics and 
Astronomy at Rutgers University where part of the work was completed and funding
by ETSF (INFRA-2007-211956).
KH acknowledges the American Chemical Society Petroleum Research Fund (48802-DNI1) 
for partial support of this research.
GK was supported by NSF under grant No. DMR-0906943.

\begin{appendix}

\section{Details of the supercell approach}
\label{app:supercell}

\begin{figure}
  \includegraphics[width=0.75\linewidth]{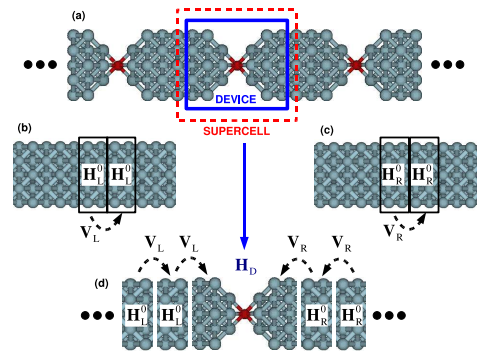}
  \caption{
    Illustration of the supercell approach to calculate the electronic structure 
    of the device and of the leads: (a) One-dimensional periodic system to calculate
    the electronic structure of the device region. (b,c) Infinite nanowires to
    calculate the electronic structure of the left (L) and right (R) semi-infinite 
    leads. (d) Sketch of the setup of the physical system: The device region (D) is 
    suspended between two semi-infinite leads L and R.
  }
  \label{fig:supercell}
\end{figure}

In order to generate the effective one-body Hamiltonians of the
device and leads the supercell approach is used. 
The electronic structure of the device region is calculated with the 
{\sc Crystal06} {\it ab initio} \cite{Crystal:06} electronic structure program for periodic 
systems by definining a one-dimensional periodic system consisting 
of the device region as the unit cell, as shown in Fig. \ref{fig:supercell}(a). 
It is crucial that the device part D contains a sufficient part of 
the nanowire electrodes so that the two leads L and R 
are far enough away from the scattering region, and the electronic 
structure in the leads has relaxed to that of a bulk (i.e. infinite) 
nanowire. In that case the electronic structure of the periodic system 
build from a periodic repetition of the device region is the same as
the electronic structure of the device between two semi-infinite
nanowires as can be seen from Fig. \ref{fig:supercell}.
Thus the device Hamiltonian $\mathbf{H}_{\rm D}$ can be obtained from 
the converged KS Hamiltonian of the unit cell of the periodic system. 

In the same way the unit cell Hamiltonians $\mathbf{H}^0_{\rm L/R}$ and 
hoppings $\mathbf{V}_{\rm L/R}$ between unit cells of the left and right 
leads are extracted from periodic calculations of infinite nanowires of 
finite width (see Fig. \ref{fig:supercell}(b,c)). Again it is crucial that the device region 
contains enough bulk electrode material so that the electronic structure 
in the electrodes is that of bulk nanowires. The lead self-energies 
$\mathbf\Sigma_{\rm L}$, $\mathbf\Sigma_{\rm R}$ which describe the coupling of the device region D
to the semi-infinite nanowires L and R in the situation depicted in Fig. \ref{fig:supercell}(d)
can now be calculated by the following Dyson equation:
\begin{eqnarray}
  \label{eq:Dyson}
  \lefteqn{\mathbf\Sigma_{\rm L/R}(\omega) =
  (\mathbf{V}_{\rm L/R}-\omega\mathbf{S}_{\rm L/R}) \times} 
  \\
  && \times(\omega\mathbf{S}^0_{\rm L/R}-\mathbf{H}^0_{\rm L/R}-\mathbf\Sigma_{\rm L/R}(\omega))^{-1}
  \,(\mathbf{V}_{\rm L/R}^\dagger-\omega\mathbf{S}_{\rm L/R}^\dagger)
  \nonumber
\end{eqnarray}
where $\mathbf{S}^0_{\rm L/R}$ and $\mathbf{S}_{\rm L/R}$ are the overlap matrices
taking into account the non-orthogonality of the basis set
within the unit cell and between unit cells, respectively.

By this procedure we have connected the device region D with two {\it semi-infinite}
nanowires that have the electronic structure of bulk i.e. {\it infinite} nanowires. 
The supercell approach and the so-called partitioning technique used here to obtain
the Green's function of a part of a system are discussed in more detail in the literature
(See e.g. refs. \onlinecite{Palacios:prb:01,Jacob:thesis:07,Pastawski:rmf:01,Taylor:prb:01A}).

\section{The NCA and OCA impurity solvers}
\label{app:nca+oca}

\begin{figure}
  \includegraphics[width=0.75\linewidth]{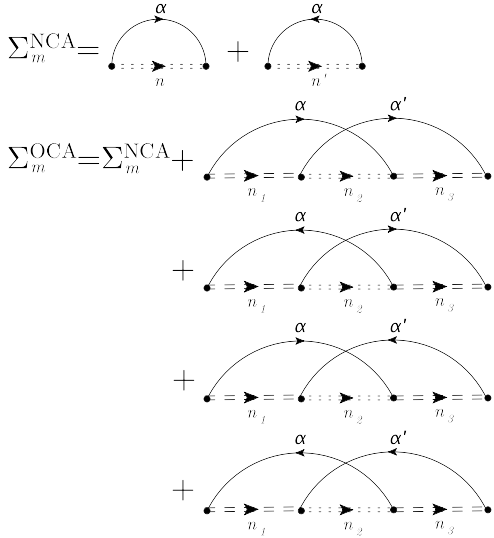}
  \caption{
    Diagrams for pseudo-particle self-energies in the NCA (first row)
    and OCA (second and following rows) for some pseudo-particle $m$. 
  }
  \label{fig:nca+oca}
\end{figure}

The general multi-orbital Anderson impurity model can be written in the
following form:
\begin{eqnarray}
  \label{eq:AIM}
  \hat{\mathcal{H}} &=& 
  \sum_{\alpha\beta} \epsilon_{\alpha}\,\hat{d}_{\alpha}^\dagger \hat{d}_{\alpha }
  + \frac{1}{2} \sum_{\alpha\beta\gamma\delta} U_{\alpha\beta\gamma\delta}
  \hat{d}_{\alpha}^\dagger \hat{d}_{\beta}^\dagger \hat{d}_{\gamma} \hat{d}_{\delta}
  \\
  &+& \sum_{k\nu\alpha} (V_{k\nu\alpha} \hat{c}_{k\nu}^\dagger \hat{d}_{\alpha}
  + V_{k\nu\alpha}^\ast \hat{d}_{\alpha}^\dagger \hat{c}_{k\nu})
  + \sum_{k\nu} \epsilon_{k\nu} \hat{c}_{k\nu}^\dagger \hat{c}_{k\nu}
  \nonumber
\end{eqnarray}
where in order to keep the notation simple we have combined the spin- and
orbital degrees of freedom into one index for each impurity level $\alpha$ 
and each band $\nu$. 

The NCA and the OCA both solve the Anderson impurity model 
by expansion in the hybridization strength around the atomic 
limit. 
The starting point is an exact diagonalization of the impurity subspace
(for example the $3d$-shell of a magnetic atom)
including the Hubbard-type interaction term:
\begin{eqnarray}
  \hat{\mathcal H}_{d} &\equiv& 
  \sum_{\alpha\beta} \epsilon_{\alpha}\,\hat{d}_{\alpha}^\dagger \hat{d}_{\alpha }
  + \frac{1}{2} \sum_{\alpha\beta\gamma\delta} U_{\alpha\beta\gamma\delta}
  \hat{d}_{\alpha}^\dagger \hat{d}_{\beta}^\dagger \hat{d}_{\gamma} \hat{d}_{\delta}
  \nonumber\\
  &\stackrel{\rm diag.}{\longrightarrow}& \sum_m \ket{m} E_m \bra{m} 
\end{eqnarray}
where $\ket{m}$ are the many-body eigenstates of $\hat{\mathcal H}_{d}$ 
and $E_m$ the respective eigen energies. 

One now introduces auxiliary fields $\hat{a}_m$, $\hat{a}_m^\dagger$ (called pseudo-particles) 
such that each impurity state is represented by a corresponding pseudo-particle:
\begin{equation}
  \hat{a}_m^\dagger\ket{\rm PPV} \equiv \ket{m}
\end{equation}
where $\ket{\rm PPV}$ is the pseudo-particle vacuum. 
The completeness of the impurity eigenstates imposes the following constraint: 
\begin{equation}
  Q \equiv \sum_m \hat{a}_m^\dagger \hat{a}_m = 1
\end{equation}
The physical electron operators $\hat{d}_{\alpha}^\dagger$
can now be expressed by the pseudo-particle (PP) operators:
\begin{equation}
  \hat{d}^\dagger_{\alpha} = \sum_{n,m} (F^{\alpha\dagger})_{nm}  \hat{a}_n^\dagger \hat{a}_m
\end{equation}
where $(F^{\alpha\dagger})_{nm}\equiv\bra{n}\hat{d}^\dagger_{\alpha}\ket{m}$ are 
the matrix elements of the impurity-electron creation operator. 
For later convenience we also define the corresponding matrix elements
of the impurity-electron destruction operator as: 
$(F^{\alpha})_{nm}\equiv\bra{n}\hat{d}_{\alpha}\ket{m}$.
The anti-commutation rules for the physical electron operators then require that
the PP $\hat{a}_m$ is a boson (fermion) if the corresponding state $\ket{m}$ 
contains an even (odd) number of electrons.

In the PP representation we can now rewrite the Hamiltonian of the
generalized Anderson impurity model as follows:
\begin{eqnarray}
  \label{eq:PP-AIM}
  \hat{\mathcal H} &=& \sum_m E_m \hat{a}_m^\dagger \hat{a}_m 
  + \sum_{k\nu} \epsilon_{k\nu} \hat{c}_{k\nu}^\dagger\hat{c}_{k\nu}  +\lambda(Q-1)
  \nonumber\\
  &+& \sum_{{mn}\atop{k\nu\alpha}}
  \left(V_{k\nu,\alpha} \hat{c}_{k\nu}^\dagger \hat{a}_m^\dagger  (F^{\alpha})_{nm} \hat{a}_n + H.c. \right)  
\end{eqnarray}
where we have included the constraint $Q\equiv 1$ into the Hamiltonian.
The corresponding Lagrange multiplier $\lambda$ can be interpreted as a 
(negative) chemical potential for the PPs.

In the PP picture, the hybridization with the bath electrons
given by the last term in eq. (\ref{eq:PP-AIM}) becomes now the interaction 
for the PPs. Because of the fermionic and bosonic commutation rules for the 
PPs, one can now develop a diagrammatic perturbation expansion in the PP 
interaction. The PP propagators can be written as
\begin{equation}
  G_m(\omega) = (\omega-\lambda-E_m-\Sigma_m(\omega))^{-1}
\end{equation}
where $\Sigma_m$ is the PP self-energy describing the dynamic interaction of the PP $m$ 
with the other PPs.

The NCA consists in taking into account the diagrams shown in the first row of 
Fig. \ref{fig:nca+oca} for some PP $m$. 
The NCA diagrams describe processes where a single electron (hole) jumps from 
the bath to the impurity and back thereby temporarily creating a PP with N+1 
(N-1) electrons. The NCA equations correspond to a self-consistent perturbation 
expansion to lowest order in the hybridization function 
$\Delta_\alpha(\omega)\equiv\sum_{k,\nu}V_{k\nu,\alpha}^\ast  V_{k\nu,\alpha}$.
Since the fermionic self-energies depend on the dressed bosonic propagators, 
and vice versa, the NCA equations have to be solved self-consistently.
Once the NCA equations are solved the physical quantities can be calculated from the PP
self-energies. 

The OCA takes into account second order diagrams where two bath electron lines cross 
as shown in the last four rows of Fig. \ref{fig:nca+oca}.
The self-energies for the PPs again depend on the full propagators of other PPs, 
and hence the OCA equations also have to be solved self-consistently.
Further details of the NCA and OCA impurity solver can be found e.g. in Refs. 
\onlinecite{Kotliar:rmp:06,Haule:prb:10,Hewson:book,Haule:prb:01,Pruschke:zphysb:89,Bickers:rmp:87}.

\end{appendix}


\bibliographystyle{apsrev}
\bibliography{matcon,footnotes}

\begin{thebibliography}{56}
\expandafter\ifx\csname natexlab\endcsname\relax\def\natexlab#1{#1}\fi
\expandafter\ifx\csname bibnamefont\endcsname\relax
  \def\bibnamefont#1{#1}\fi
\expandafter\ifx\csname bibfnamefont\endcsname\relax
  \def\bibfnamefont#1{#1}\fi
\expandafter\ifx\csname citenamefont\endcsname\relax
  \def\citenamefont#1{#1}\fi
\expandafter\ifx\csname url\endcsname\relax
  \def\url#1{\texttt{#1}}\fi
\expandafter\ifx\csname urlprefix\endcsname\relax\def\urlprefix{URL }\fi
\providecommand{\bibinfo}[2]{#2}
\providecommand{\eprint}[2][]{\url{#2}}

\bibitem[{\citenamefont{Palacios et~al.}(2001)\citenamefont{Palacios,
  P\'erez-Jim\'enez, Louis, and Verg\'es}}]{Palacios:prb:01}
\bibinfo{author}{\bibfnamefont{J.~J.} \bibnamefont{Palacios}},
  \bibinfo{author}{\bibfnamefont{A.~J.} \bibnamefont{P\'erez-Jim\'enez}},
  \bibinfo{author}{\bibfnamefont{E.}~\bibnamefont{Louis}}, \bibnamefont{and}
  \bibinfo{author}{\bibfnamefont{J.~A.} \bibnamefont{Verg\'es}},
  \bibinfo{journal}{Phys.\ Rev.\ B} \textbf{\bibinfo{volume}{64}},
  \bibinfo{pages}{115411} (\bibinfo{year}{2001}).

\bibitem[{\citenamefont{Taylor et~al.}(2001)\citenamefont{Taylor, Guo, and
  Wang}}]{Taylor:prb:01A}
\bibinfo{author}{\bibfnamefont{J.}~\bibnamefont{Taylor}},
  \bibinfo{author}{\bibfnamefont{H.}~\bibnamefont{Guo}}, \bibnamefont{and}
  \bibinfo{author}{\bibfnamefont{J.}~\bibnamefont{Wang}},
  \bibinfo{journal}{Phys. Rev. B} \textbf{\bibinfo{volume}{63}},
  \bibinfo{pages}{121104} (\bibinfo{year}{2001}).

\bibitem[{\citenamefont{Agra\"{\i}t et~al.}(2003)\citenamefont{Agra\"{\i}t,
  Yeyati, and van Ruitenbeek}}]{Agrait:pr:03}
\bibinfo{author}{\bibfnamefont{N.}~\bibnamefont{Agra\"{\i}t}},
  \bibinfo{author}{\bibfnamefont{A.~L.} \bibnamefont{Yeyati}},
  \bibnamefont{and} \bibinfo{author}{\bibfnamefont{J.~M.} \bibnamefont{van
  Ruitenbeek}}, \bibinfo{journal}{Physics Reports}
  \textbf{\bibinfo{volume}{377}}, \bibinfo{pages}{81} (\bibinfo{year}{2003}),
  \bibinfo{note}{and references therein}.

\bibitem[{\citenamefont{Viret et~al.}(2002)\citenamefont{Viret, Berger,
  Gabureac, Ott, Olligs, Petej, Gregg, Fermon, Francinet, and
  LeGoff}}]{Viret:prb:02}
\bibinfo{author}{\bibfnamefont{M.}~\bibnamefont{Viret}},
  \bibinfo{author}{\bibfnamefont{S.}~\bibnamefont{Berger}},
  \bibinfo{author}{\bibfnamefont{M.}~\bibnamefont{Gabureac}},
  \bibinfo{author}{\bibfnamefont{F.}~\bibnamefont{Ott}},
  \bibinfo{author}{\bibfnamefont{D.}~\bibnamefont{Olligs}},
  \bibinfo{author}{\bibfnamefont{I.}~\bibnamefont{Petej}},
  \bibinfo{author}{\bibfnamefont{J.~F.} \bibnamefont{Gregg}},
  \bibinfo{author}{\bibfnamefont{C.}~\bibnamefont{Fermon}},
  \bibinfo{author}{\bibfnamefont{G.}~\bibnamefont{Francinet}},
  \bibnamefont{and} \bibinfo{author}{\bibfnamefont{G.}~\bibnamefont{LeGoff}},
  \bibinfo{journal}{Phys.\ Rev.\ B} \textbf{\bibinfo{volume}{66}},
  \bibinfo{pages}{220401(R)} (\bibinfo{year}{2002}).

\bibitem[{\citenamefont{Untiedt et~al.}(2004)\citenamefont{Untiedt, Dekker,
  Djukic, , and van Ruitenbeek}}]{Untiedt:prb:04}
\bibinfo{author}{\bibfnamefont{C.}~\bibnamefont{Untiedt}},
  \bibinfo{author}{\bibfnamefont{D.~M.~T.} \bibnamefont{Dekker}},
  \bibinfo{author}{\bibfnamefont{D.}~\bibnamefont{Djukic}}, , \bibnamefont{and}
  \bibinfo{author}{\bibfnamefont{J.~M.} \bibnamefont{van Ruitenbeek}},
  \bibinfo{journal}{Phys.\ Rev.\ B} \textbf{\bibinfo{volume}{69}},
  \bibinfo{pages}{081401(R)} (\bibinfo{year}{2004}).

\bibitem[{\citenamefont{Keane et~al.}(2006)\citenamefont{Keane, Yu, and
  Natelson}}]{Keane:apl:06}
\bibinfo{author}{\bibfnamefont{Z.~K.} \bibnamefont{Keane}},
  \bibinfo{author}{\bibfnamefont{L.~H.} \bibnamefont{Yu}}, \bibnamefont{and}
  \bibinfo{author}{\bibfnamefont{D.}~\bibnamefont{Natelson}},
  \bibinfo{journal}{Appl. Phys. Lett.} \textbf{\bibinfo{volume}{88}},
  \bibinfo{pages}{062514} (\bibinfo{year}{2006}).

\bibitem[{\citenamefont{Bolotin et~al.}(2006)\citenamefont{Bolotin, Kuemmeth,
  Pasupathy, and Ralph}}]{Bolotin:nl:06}
\bibinfo{author}{\bibfnamefont{K.~I.} \bibnamefont{Bolotin}},
  \bibinfo{author}{\bibfnamefont{F.}~\bibnamefont{Kuemmeth}},
  \bibinfo{author}{\bibfnamefont{A.~N.} \bibnamefont{Pasupathy}},
  \bibnamefont{and} \bibinfo{author}{\bibfnamefont{D.~C.} \bibnamefont{Ralph}},
  \bibinfo{journal}{Nano Lett.} \textbf{\bibinfo{volume}{6}},
  \bibinfo{pages}{123} (\bibinfo{year}{2006}).

\bibitem[{\citenamefont{Jacob et~al.}(2005)\citenamefont{Jacob,
  Fern\'andez-Rossier, and Palacios}}]{Jacob:prb:05}
\bibinfo{author}{\bibfnamefont{D.}~\bibnamefont{Jacob}},
  \bibinfo{author}{\bibfnamefont{J.}~\bibnamefont{Fern\'andez-Rossier}},
  \bibnamefont{and} \bibinfo{author}{\bibfnamefont{J.~J.}
  \bibnamefont{Palacios}}, \bibinfo{journal}{Phys. Rev. B}
  \textbf{\bibinfo{volume}{71}}, \bibinfo{pages}{220403(R)}
  (\bibinfo{year}{2005}).

\bibitem[{\citenamefont{Smogunov et~al.}(2006)\citenamefont{Smogunov,
  Dal{}Corso, and Tosatti}}]{Smogunov:prb:06}
\bibinfo{author}{\bibfnamefont{A.}~\bibnamefont{Smogunov}},
  \bibinfo{author}{\bibfnamefont{A.}~\bibnamefont{Dal{}Corso}},
  \bibnamefont{and} \bibinfo{author}{\bibfnamefont{E.}~\bibnamefont{Tosatti}},
  \bibinfo{journal}{Phys. Rev. B} \textbf{\bibinfo{volume}{73}},
  \bibinfo{pages}{075418} (\bibinfo{year}{2006}).

\bibitem[{\citenamefont{Kondo}(1964)}]{Kondo:ptp:64}
\bibinfo{author}{\bibfnamefont{J.}~\bibnamefont{Kondo}},
  \bibinfo{journal}{Prog. Theor. Phys.} \textbf{\bibinfo{volume}{32}},
  \bibinfo{pages}{37} (\bibinfo{year}{1964}).

\bibitem[{\citenamefont{Calvo et~al.}(2009)\citenamefont{Calvo,
  Fern\'andez-Rossier, Palacios, Jacob, Natelson, and
  Untiedt}}]{Calvo:nature:09}
\bibinfo{author}{\bibfnamefont{M.~R.} \bibnamefont{Calvo}},
  \bibinfo{author}{\bibfnamefont{J.}~\bibnamefont{Fern\'andez-Rossier}},
  \bibinfo{author}{\bibfnamefont{J.~J.} \bibnamefont{Palacios}},
  \bibinfo{author}{\bibfnamefont{D.}~\bibnamefont{Jacob}},
  \bibinfo{author}{\bibfnamefont{D.}~\bibnamefont{Natelson}}, \bibnamefont{and}
  \bibinfo{author}{\bibfnamefont{C.}~\bibnamefont{Untiedt}},
  \bibinfo{journal}{Nature} \textbf{\bibinfo{volume}{358}},
  \bibinfo{pages}{1150} (\bibinfo{year}{2009}).

\bibitem[{\citenamefont{Fano}(1961)}]{Fano:pr:61}
\bibinfo{author}{\bibfnamefont{U.}~\bibnamefont{Fano}}, \bibinfo{journal}{Phys.
  Rev.} \textbf{\bibinfo{volume}{124}}, \bibinfo{pages}{1866}
  (\bibinfo{year}{1961}).

\bibitem[{\citenamefont{Madhavan et~al.}(1998)\citenamefont{Madhavan, Chen,
  Jamneala, Crommie, and Wingreen}}]{Madhavan:science:98}
\bibinfo{author}{\bibfnamefont{V.}~\bibnamefont{Madhavan}},
  \bibinfo{author}{\bibfnamefont{W.}~\bibnamefont{Chen}},
  \bibinfo{author}{\bibfnamefont{T.}~\bibnamefont{Jamneala}},
  \bibinfo{author}{\bibfnamefont{M.~F.} \bibnamefont{Crommie}},
  \bibnamefont{and} \bibinfo{author}{\bibfnamefont{N.~S.}
  \bibnamefont{Wingreen}}, \bibinfo{journal}{Science}
  \textbf{\bibinfo{volume}{280}}, \bibinfo{pages}{567} (\bibinfo{year}{1998}).

\bibitem[{\citenamefont{Vitali et~al.}(2008)\citenamefont{Vitali, Ohmann,
  Stepanow, Gambardella, Tao, Huang, Stepanyuk, Bruno, and
  Kern}}]{Vitali:prl:08}
\bibinfo{author}{\bibfnamefont{L.}~\bibnamefont{Vitali}},
  \bibinfo{author}{\bibfnamefont{R.}~\bibnamefont{Ohmann}},
  \bibinfo{author}{\bibfnamefont{S.}~\bibnamefont{Stepanow}},
  \bibinfo{author}{\bibfnamefont{P.}~\bibnamefont{Gambardella}},
  \bibinfo{author}{\bibfnamefont{K.}~\bibnamefont{Tao}},
  \bibinfo{author}{\bibfnamefont{R.}~\bibnamefont{Huang}},
  \bibinfo{author}{\bibfnamefont{V.~S.} \bibnamefont{Stepanyuk}},
  \bibinfo{author}{\bibfnamefont{P.}~\bibnamefont{Bruno}}, \bibnamefont{and}
  \bibinfo{author}{\bibfnamefont{K.}~\bibnamefont{Kern}},
  \bibinfo{journal}{Phys. Rev. Lett.} \textbf{\bibinfo{volume}{101}},
  \bibinfo{pages}{216802} (\bibinfo{year}{2008}).

\bibitem[{\citenamefont{Ne\'el et~al.}(2008)\citenamefont{Ne\'el, Kr\"oger,
  Berndt, Wehling, Lichtenstein, and Katsnelson}}]{Neel:prl:08}
\bibinfo{author}{\bibfnamefont{N.}~\bibnamefont{Ne\'el}},
  \bibinfo{author}{\bibfnamefont{J.}~\bibnamefont{Kr\"oger}},
  \bibinfo{author}{\bibfnamefont{R.}~\bibnamefont{Berndt}},
  \bibinfo{author}{\bibfnamefont{T.}~\bibnamefont{Wehling}},
  \bibinfo{author}{\bibfnamefont{A.}~\bibnamefont{Lichtenstein}},
  \bibnamefont{and} \bibinfo{author}{\bibfnamefont{M.~I.}
  \bibnamefont{Katsnelson}}, \bibinfo{journal}{Phys. Rev. Lett.}
  \textbf{\bibinfo{volume}{101}}, \bibinfo{pages}{266803}
  (\bibinfo{year}{2008}).

\bibitem[{\citenamefont{N\'eel et~al.}(2007)\citenamefont{N\'eel, Kr\"oger,
  Limot, Palotas, Hofer, and Berndt}}]{Neel:prl:07}
\bibinfo{author}{\bibfnamefont{N.}~\bibnamefont{N\'eel}},
  \bibinfo{author}{\bibfnamefont{J.}~\bibnamefont{Kr\"oger}},
  \bibinfo{author}{\bibfnamefont{L.}~\bibnamefont{Limot}},
  \bibinfo{author}{\bibfnamefont{K.}~\bibnamefont{Palotas}},
  \bibinfo{author}{\bibfnamefont{W.~A.} \bibnamefont{Hofer}}, \bibnamefont{and}
  \bibinfo{author}{\bibfnamefont{R.}~\bibnamefont{Berndt}},
  \bibinfo{journal}{Phys. Rev. Lett.} \textbf{\bibinfo{volume}{98}},
  \bibinfo{pages}{016801} (\bibinfo{year}{2007}).

\bibitem[{\citenamefont{Hewson}(1993)}]{Hewson:book}
\bibinfo{author}{\bibfnamefont{A.~C.} \bibnamefont{Hewson}},
  \emph{\bibinfo{title}{The Kondo problem to heavy fermions}}
  (\bibinfo{publisher}{Cambridge University Press}, \bibinfo{year}{1993}).

\bibitem[{\citenamefont{Thygesen and Rubio}(2007)}]{Thygesen:jcp:07}
\bibinfo{author}{\bibfnamefont{K.~S.} \bibnamefont{Thygesen}} \bibnamefont{and}
  \bibinfo{author}{\bibfnamefont{A.}~\bibnamefont{Rubio}}, \bibinfo{journal}{J
  . Chem. Phys.} \textbf{\bibinfo{volume}{126}}, \bibinfo{pages}{091101}
  (\bibinfo{year}{2007}).

\bibitem[{\citenamefont{Ferretti et~al.}(2005)\citenamefont{Ferretti,
  Calzolari, Di~Felice, Manghi, Caldas, Buongiorno~Nardelli, and
  Molinari}}]{Ferretti:prl:05}
\bibinfo{author}{\bibfnamefont{A.}~\bibnamefont{Ferretti}},
  \bibinfo{author}{\bibfnamefont{A.}~\bibnamefont{Calzolari}},
  \bibinfo{author}{\bibfnamefont{R.}~\bibnamefont{Di~Felice}},
  \bibinfo{author}{\bibfnamefont{F.}~\bibnamefont{Manghi}},
  \bibinfo{author}{\bibfnamefont{M.~J.} \bibnamefont{Caldas}},
  \bibinfo{author}{\bibfnamefont{M.}~\bibnamefont{Buongiorno~Nardelli}},
  \bibnamefont{and} \bibinfo{author}{\bibfnamefont{E.}~\bibnamefont{Molinari}},
  \bibinfo{journal}{Phys. Rev. Lett.} \textbf{\bibinfo{volume}{94}},
  \bibinfo{pages}{116802} (\bibinfo{year}{2005}).

\bibitem[{\citenamefont{Jacob et~al.}(2009)\citenamefont{Jacob, Haule, and
  Kotliar}}]{Jacob:prl:09}
\bibinfo{author}{\bibfnamefont{D.}~\bibnamefont{Jacob}},
  \bibinfo{author}{\bibfnamefont{K.}~\bibnamefont{Haule}}, \bibnamefont{and}
  \bibinfo{author}{\bibfnamefont{G.}~\bibnamefont{Kotliar}},
  \bibinfo{journal}{Phys. Rev. Lett.} \textbf{\bibinfo{volume}{103}},
  \bibinfo{pages}{016803} (\bibinfo{year}{2009}).

\bibitem[{\citenamefont{Lucignano et~al.}(2009)\citenamefont{Lucignano,
  Mazzarello, Smogunov, Fabrizio, and Tosatti}}]{Lucignano:natmat:09}
\bibinfo{author}{\bibfnamefont{P.}~\bibnamefont{Lucignano}},
  \bibinfo{author}{\bibfnamefont{R.}~\bibnamefont{Mazzarello}},
  \bibinfo{author}{\bibfnamefont{A.}~\bibnamefont{Smogunov}},
  \bibinfo{author}{\bibfnamefont{M.}~\bibnamefont{Fabrizio}}, \bibnamefont{and}
  \bibinfo{author}{\bibfnamefont{E.}~\bibnamefont{Tosatti}},
  \bibinfo{journal}{Nature Materials} \textbf{\bibinfo{volume}{8}},
  \bibinfo{pages}{563} (\bibinfo{year}{2009}).

\bibitem[{\citenamefont{Dias~da Silva et~al.}(2009)\citenamefont{Dias~da Silva,
  Tiago, Ulloa, Reboredo, and Dagotto}}]{DiasdaSilva:prb:09}
\bibinfo{author}{\bibfnamefont{L.~G.~G.~V.} \bibnamefont{Dias~da Silva}},
  \bibinfo{author}{\bibfnamefont{M.~L.} \bibnamefont{Tiago}},
  \bibinfo{author}{\bibfnamefont{S.~E.} \bibnamefont{Ulloa}},
  \bibinfo{author}{\bibfnamefont{F.~A.} \bibnamefont{Reboredo}},
  \bibnamefont{and} \bibinfo{author}{\bibfnamefont{E.}~\bibnamefont{Dagotto}},
  \bibinfo{journal}{Phys. Rev. B} \textbf{\bibinfo{volume}{80}},
  \bibinfo{pages}{155443} (\bibinfo{year}{2009}).

\bibitem[{\citenamefont{Jacob and Kotliar}(2010)}]{Jacob:prb:10}
\bibinfo{author}{\bibfnamefont{D.}~\bibnamefont{Jacob}} \bibnamefont{and}
  \bibinfo{author}{\bibfnamefont{G.}~\bibnamefont{Kotliar}},
  \bibinfo{journal}{Phys. Rev. B} \textbf{\bibinfo{volume}{82}},
  \bibinfo{pages}{085423} (\bibinfo{year}{2010}).

\bibitem[{\citenamefont{Georges et~al.}(1996)\citenamefont{Georges, Kotliar,
  Krauth, and Rozenberg}}]{Georges:rmp:96}
\bibinfo{author}{\bibfnamefont{A.}~\bibnamefont{Georges}},
  \bibinfo{author}{\bibfnamefont{G.}~\bibnamefont{Kotliar}},
  \bibinfo{author}{\bibfnamefont{W.}~\bibnamefont{Krauth}}, \bibnamefont{and}
  \bibinfo{author}{\bibfnamefont{M.~J.} \bibnamefont{Rozenberg}},
  \bibinfo{journal}{Rev. Mod. Phys.} \textbf{\bibinfo{volume}{68}},
  \bibinfo{pages}{13} (\bibinfo{year}{1996}).

\bibitem[{\citenamefont{Vollhardt et~al.}(1997)\citenamefont{Vollhardt,
  Bl\"umer, Held, Kollar, Schlipf, and Ulmke}}]{Vollhardt:zpb:97}
\bibinfo{author}{\bibfnamefont{D.}~\bibnamefont{Vollhardt}},
  \bibinfo{author}{\bibfnamefont{N.}~\bibnamefont{Bl\"umer}},
  \bibinfo{author}{\bibfnamefont{K.}~\bibnamefont{Held}},
  \bibinfo{author}{\bibfnamefont{M.}~\bibnamefont{Kollar}},
  \bibinfo{author}{\bibfnamefont{J.}~\bibnamefont{Schlipf}}, \bibnamefont{and}
  \bibinfo{author}{\bibfnamefont{M.}~\bibnamefont{Ulmke}},
  \bibinfo{journal}{Zeit. f. Phys. B} \textbf{\bibinfo{volume}{103}},
  \bibinfo{pages}{283} (\bibinfo{year}{1997}).

\bibitem[{\citenamefont{Dobrosavljevic and
  Kotliar}(1998)}]{Dobrosavljevic:philtrans:98}
\bibinfo{author}{\bibfnamefont{V.}~\bibnamefont{Dobrosavljevic}}
  \bibnamefont{and} \bibinfo{author}{\bibfnamefont{G.}~\bibnamefont{Kotliar}},
  \bibinfo{journal}{Philos. Trans. R. Soc. London, Ser. A}
  \textbf{\bibinfo{volume}{356}}, \bibinfo{pages}{57} (\bibinfo{year}{1998}).

\bibitem[{\citenamefont{Potthoff and Nolting}(1999)}]{Potthoff:prb:99}
\bibinfo{author}{\bibfnamefont{M.}~\bibnamefont{Potthoff}} \bibnamefont{and}
  \bibinfo{author}{\bibfnamefont{W.}~\bibnamefont{Nolting}},
  \bibinfo{journal}{Phys. Rev. B} \textbf{\bibinfo{volume}{60}},
  \bibinfo{pages}{7834} (\bibinfo{year}{1999}).

\bibitem[{\citenamefont{Okamoto and Millis}(2004)}]{Okamoto:nature:04}
\bibinfo{author}{\bibfnamefont{S.}~\bibnamefont{Okamoto}} \bibnamefont{and}
  \bibinfo{author}{\bibfnamefont{A.~J.} \bibnamefont{Millis}},
  \bibinfo{journal}{Nature} \textbf{\bibinfo{volume}{428}},
  \bibinfo{pages}{630} (\bibinfo{year}{2004}).

\bibitem[{\citenamefont{Freericks}(2006)}]{Freericks:book:06}
\bibinfo{author}{\bibfnamefont{J.~K.} \bibnamefont{Freericks}},
  \emph{\bibinfo{title}{Transport in Multilayered Nanostructures: The Dynamical
  Mean-Field Theory Approach}} (\bibinfo{publisher}{Imperial College Press},
  \bibinfo{address}{London}, \bibinfo{year}{2006}).

\bibitem[{\citenamefont{Gorelik et~al.}(2010)\citenamefont{Gorelik, Titvinidze,
  Hofstetter, Snoek, and Bl\"umer}}]{Gorelik:10}
\bibinfo{author}{\bibfnamefont{E.~V.} \bibnamefont{Gorelik}},
  \bibinfo{author}{\bibfnamefont{I.}~\bibnamefont{Titvinidze}},
  \bibinfo{author}{\bibfnamefont{W.}~\bibnamefont{Hofstetter}},
  \bibinfo{author}{\bibfnamefont{M.}~\bibnamefont{Snoek}}, \bibnamefont{and}
  \bibinfo{author}{\bibfnamefont{N.}~\bibnamefont{Bl\"umer}},
  \bibinfo{journal}{Phys. Rev. Lett.} \textbf{\bibinfo{volume}{105}},
  \bibinfo{pages}{065301} (\bibinfo{year}{2010}).

\bibitem[{\citenamefont{Kotliar et~al.}(2006)\citenamefont{Kotliar, Savrasov,
  Haule, Oudovenko, Parcollet, and Marianetti}}]{Kotliar:rmp:06}
\bibinfo{author}{\bibfnamefont{G.}~\bibnamefont{Kotliar}},
  \bibinfo{author}{\bibfnamefont{S.~Y.} \bibnamefont{Savrasov}},
  \bibinfo{author}{\bibfnamefont{K.}~\bibnamefont{Haule}},
  \bibinfo{author}{\bibfnamefont{V.~S.} \bibnamefont{Oudovenko}},
  \bibinfo{author}{\bibfnamefont{O.}~\bibnamefont{Parcollet}},
  \bibnamefont{and} \bibinfo{author}{\bibfnamefont{C.~A.}
  \bibnamefont{Marianetti}}, \bibinfo{journal}{Rev. Mod. Phys.}
  \textbf{\bibinfo{volume}{78}}, \bibinfo{pages}{865} (\bibinfo{year}{2006}).

\bibitem[{\citenamefont{Held}(2007)}]{Held:advphys:07}
\bibinfo{author}{\bibfnamefont{K.}~\bibnamefont{Held}}, \bibinfo{journal}{Adv.
  Phys.} \textbf{\bibinfo{volume}{56}}, \bibinfo{pages}{829}
  (\bibinfo{year}{2007}).

\bibitem[{\citenamefont{Jacob}(2007)}]{Jacob:thesis:07}
\bibinfo{author}{\bibfnamefont{D.}~\bibnamefont{Jacob}}, Ph.D. thesis,
  \bibinfo{school}{Universidad de Alicante} (\bibinfo{year}{2007}),
  \bibinfo{note}{arXiv:0712.1383v1}.

\bibitem[{\citenamefont{Valli et~al.}(2010)\citenamefont{Valli, Sangiovanni,
  Gunnarsson, Toschi, and Held}}]{Valli:prl:10}
\bibinfo{author}{\bibfnamefont{A.}~\bibnamefont{Valli}},
  \bibinfo{author}{\bibfnamefont{G.}~\bibnamefont{Sangiovanni}},
  \bibinfo{author}{\bibfnamefont{O.}~\bibnamefont{Gunnarsson}},
  \bibinfo{author}{\bibfnamefont{A.}~\bibnamefont{Toschi}}, \bibnamefont{and}
  \bibinfo{author}{\bibfnamefont{K.}~\bibnamefont{Held}},
  \bibinfo{journal}{Phys. Rev. Lett.} \textbf{\bibinfo{volume}{104}},
  \bibinfo{pages}{246402} (\bibinfo{year}{2010}).

\bibitem[{\citenamefont{Florens}(2007)}]{Florens:prl:07}
\bibinfo{author}{\bibfnamefont{S.}~\bibnamefont{Florens}},
  \bibinfo{journal}{Phys. Rev. Lett.} \textbf{\bibinfo{volume}{99}},
  \bibinfo{pages}{046402} (\bibinfo{year}{2007}).

\bibitem[{\citenamefont{Dovesi et~al.}()\citenamefont{Dovesi, Saunders, Roetti,
  Orlando, Zicovich-Wilson, Pascale, Civalleri, Doll, Harrison, Bush
  et~al.}}]{Crystal:06}
\bibinfo{author}{\bibfnamefont{R.}~\bibnamefont{Dovesi}},
  \bibinfo{author}{\bibfnamefont{V.}~\bibnamefont{Saunders}},
  \bibinfo{author}{\bibfnamefont{C.}~\bibnamefont{Roetti}},
  \bibinfo{author}{\bibfnamefont{R.}~\bibnamefont{Orlando}},
  \bibinfo{author}{\bibfnamefont{C.}~\bibnamefont{Zicovich-Wilson}},
  \bibinfo{author}{\bibfnamefont{F.}~\bibnamefont{Pascale}},
  \bibinfo{author}{\bibfnamefont{B.}~\bibnamefont{Civalleri}},
  \bibinfo{author}{\bibfnamefont{K.}~\bibnamefont{Doll}},
  \bibinfo{author}{\bibfnamefont{N.}~\bibnamefont{Harrison}},
  \bibinfo{author}{\bibfnamefont{I.~J.} \bibnamefont{Bush}},
  \bibnamefont{et~al.}, \bibinfo{howpublished}{CRYSTAL06, Release 1.0.2,
  Theoretical Chemistry Group - Universita' Di Torino - Torino (Italy)}.

\bibitem[{\citenamefont{Grechnev et~al.}(2007)\citenamefont{Grechnev,
  Di{}Marco, Katsnelson, Lichtenstein, Wills, and Eriksson}}]{Grechnev:prb:07}
\bibinfo{author}{\bibfnamefont{A.}~\bibnamefont{Grechnev}},
  \bibinfo{author}{\bibfnamefont{I.}~\bibnamefont{Di{}Marco}},
  \bibinfo{author}{\bibfnamefont{M.~I.} \bibnamefont{Katsnelson}},
  \bibinfo{author}{\bibfnamefont{A.~I.} \bibnamefont{Lichtenstein}},
  \bibinfo{author}{\bibfnamefont{J.}~\bibnamefont{Wills}}, \bibnamefont{and}
  \bibinfo{author}{\bibfnamefont{O.}~\bibnamefont{Eriksson}},
  \bibinfo{journal}{Phys. Rev. B} \textbf{\bibinfo{volume}{76}},
  \bibinfo{pages}{035107} (\bibinfo{year}{2007}).

\bibitem[{foo({\natexlab{a}})}]{foot2}
\bibinfo{note}{Here we use a minimal basis set plus effective core
  pseudo-potential that takes into account the $4s$, $4p$ and $3d$ valence
  shells of the magnetic atoms.\cite{Hurley:jcp:86} $\mathbf{P}_d^ {(i)}$
  projects onto the Gaussian-type orbitals with $d$-symmetry representing the
  $3d$-shell of the atom $i$.}

\bibitem[{\citenamefont{Petukhov et~al.}(2003)\citenamefont{Petukhov, Mazin,
  Chioncel, and Lichtenstein}}]{Petukhov:prb:03}
\bibinfo{author}{\bibfnamefont{A.~G.} \bibnamefont{Petukhov}},
  \bibinfo{author}{\bibfnamefont{I.~I.} \bibnamefont{Mazin}},
  \bibinfo{author}{\bibfnamefont{L.}~\bibnamefont{Chioncel}}, \bibnamefont{and}
  \bibinfo{author}{\bibfnamefont{A.~I.} \bibnamefont{Lichtenstein}},
  \bibinfo{journal}{Phys. Rev. B} \textbf{\bibinfo{volume}{67}},
  \bibinfo{pages}{153106} (\bibinfo{year}{2003}).

\bibitem[{foo({\natexlab{b}})}]{foot3}
\bibinfo{note}{Here we have orthogonalized the complete correlated subspace C
  composed of the $3d$-shells of all magnetic atoms which is treated within
  DMFT while the rest of the system is still non-orthogonal. Also in our
  treatment the correlated subspace C has overlap with the rest of the system.
  The use of non-orthogonal basis sets in the transport problem has been
  discussed e.g. in Ref. \onlinecite{Thygesen:prb:06}.}

\bibitem[{foo({\natexlab{c}})}]{foot1}
\bibinfo{note}{In the many-body and DMFT community, one would rather call
  $\Sigma_{\rm L}$ and $\Sigma_{\rm R}$ hybridization funtions since they
  describe the dynamic hybridization of the device region with semi-infinite
  leads.}

\bibitem[{\citenamefont{Bickers}(1987)}]{Bickers:rmp:87}
\bibinfo{author}{\bibfnamefont{N.~E.} \bibnamefont{Bickers}},
  \bibinfo{journal}{Rev. Mod. Phys.} \textbf{\bibinfo{volume}{59}},
  \bibinfo{pages}{845} (\bibinfo{year}{1987}).

\bibitem[{\citenamefont{Pruschke and Grewe}(1989)}]{Pruschke:zphysb:89}
\bibinfo{author}{\bibfnamefont{T.}~\bibnamefont{Pruschke}} \bibnamefont{and}
  \bibinfo{author}{\bibfnamefont{N.}~\bibnamefont{Grewe}}, \bibinfo{journal}{Z.
  Phys. B: Condens. Matter} \textbf{\bibinfo{volume}{74}}, \bibinfo{pages}{439}
  (\bibinfo{year}{1989}).

\bibitem[{\citenamefont{Haule et~al.}(2001)\citenamefont{Haule, Kirchner,
  Kroha, and W\"olfle}}]{Haule:prb:01}
\bibinfo{author}{\bibfnamefont{K.}~\bibnamefont{Haule}},
  \bibinfo{author}{\bibfnamefont{S.}~\bibnamefont{Kirchner}},
  \bibinfo{author}{\bibfnamefont{J.}~\bibnamefont{Kroha}}, \bibnamefont{and}
  \bibinfo{author}{\bibfnamefont{P.}~\bibnamefont{W\"olfle}},
  \bibinfo{journal}{Phys. Rev. B} \textbf{\bibinfo{volume}{64}},
  \bibinfo{pages}{155111} (\bibinfo{year}{2001}).

\bibitem[{\citenamefont{Haule et~al.}(2010)\citenamefont{Haule, Yee, and
  Kim}}]{Haule:prb:10}
\bibinfo{author}{\bibfnamefont{K.}~\bibnamefont{Haule}},
  \bibinfo{author}{\bibfnamefont{C.-H.} \bibnamefont{Yee}}, \bibnamefont{and}
  \bibinfo{author}{\bibfnamefont{K.}~\bibnamefont{Kim}},
  \bibinfo{journal}{Phys. Rev. B} \textbf{\bibinfo{volume}{81}},
  \bibinfo{pages}{195107} (\bibinfo{year}{2010}).

\bibitem[{\citenamefont{Meir and Wingreen}(1992)}]{Meir:prl:92}
\bibinfo{author}{\bibfnamefont{Y.}~\bibnamefont{Meir}} \bibnamefont{and}
  \bibinfo{author}{\bibfnamefont{N.~S.} \bibnamefont{Wingreen}},
  \bibinfo{journal}{Phys.\ Rev.\ Lett.} \textbf{\bibinfo{volume}{68}},
  \bibinfo{pages}{2512} (\bibinfo{year}{1992}).

\bibitem[{\citenamefont{Werner et~al.}(2010)\citenamefont{Werner, Oka,
  Eckstein, and Millis}}]{Werner:prb:10}
\bibinfo{author}{\bibfnamefont{P.}~\bibnamefont{Werner}},
  \bibinfo{author}{\bibfnamefont{T.}~\bibnamefont{Oka}},
  \bibinfo{author}{\bibfnamefont{M.}~\bibnamefont{Eckstein}}, \bibnamefont{and}
  \bibinfo{author}{\bibfnamefont{A.~J.} \bibnamefont{Millis}},
  \bibinfo{journal}{Phys. Rev. B} \textbf{\bibinfo{volume}{81}},
  \bibinfo{pages}{035108} (\bibinfo{year}{2010}).

\bibitem[{\citenamefont{Dirks et~al.}(2010)\citenamefont{Dirks, Werner,
  Jarrell, and Pruschke}}]{Dirks:pre:10}
\bibinfo{author}{\bibfnamefont{A.}~\bibnamefont{Dirks}},
  \bibinfo{author}{\bibfnamefont{P.}~\bibnamefont{Werner}},
  \bibinfo{author}{\bibfnamefont{M.}~\bibnamefont{Jarrell}}, \bibnamefont{and}
  \bibinfo{author}{\bibfnamefont{T.}~\bibnamefont{Pruschke}},
  \bibinfo{journal}{Phys. Rev. E} \textbf{\bibinfo{volume}{82}},
  \bibinfo{pages}{026701} (\bibinfo{year}{2010}).

\bibitem[{\citenamefont{Landauer}(1970)}]{Landauer:philmag:70}
\bibinfo{author}{\bibfnamefont{R.}~\bibnamefont{Landauer}},
  \bibinfo{journal}{Philos. Mag.} \textbf{\bibinfo{volume}{21}},
  \bibinfo{pages}{863} (\bibinfo{year}{1970}).

\bibitem[{\citenamefont{Jacob et~al.}(2006)\citenamefont{Jacob,
  Fern\'andez-Rossier, and Palacios}}]{Jacob:prb:06b}
\bibinfo{author}{\bibfnamefont{D.}~\bibnamefont{Jacob}},
  \bibinfo{author}{\bibfnamefont{J.}~\bibnamefont{Fern\'andez-Rossier}},
  \bibnamefont{and} \bibinfo{author}{\bibfnamefont{J.~J.}
  \bibnamefont{Palacios}}, \bibinfo{journal}{Phys. Rev. B}
  \textbf{\bibinfo{volume}{74}}, \bibinfo{pages}{081402}
  (\bibinfo{year}{2006}).

\bibitem[{\citenamefont{Jacob et~al.}(2008)\citenamefont{Jacob,
  Fern\'andez-Rossier, and Palacios}}]{Jacob:prb:08}
\bibinfo{author}{\bibfnamefont{D.}~\bibnamefont{Jacob}},
  \bibinfo{author}{\bibfnamefont{J.}~\bibnamefont{Fern\'andez-Rossier}},
  \bibnamefont{and} \bibinfo{author}{\bibfnamefont{J.~J.}
  \bibnamefont{Palacios}}, \bibinfo{journal}{Phys. Rev. B}
  \textbf{\bibinfo{volume}{77}}, \bibinfo{pages}{165412}
  (\bibinfo{year}{2008}).

\bibitem[{\citenamefont{Thygesen and Rubio}(2008)}]{Thygesen:prb:08}
\bibinfo{author}{\bibfnamefont{K.~S.} \bibnamefont{Thygesen}} \bibnamefont{and}
  \bibinfo{author}{\bibfnamefont{A.}~\bibnamefont{Rubio}},
  \bibinfo{journal}{Phys. Rev. B} \textbf{\bibinfo{volume}{77}},
  \bibinfo{pages}{115333} (\bibinfo{year}{2008}).

\bibitem[{\citenamefont{Tard et~al.}(2005)\citenamefont{Tard, Liu, Ibrahim,
  Bruschi, Gioia, Davies, Yang, Wang, Sawers, and Pickett}}]{Tard:nature:05}
\bibinfo{author}{\bibfnamefont{C.}~\bibnamefont{Tard}},
  \bibinfo{author}{\bibfnamefont{X.}~\bibnamefont{Liu}},
  \bibinfo{author}{\bibfnamefont{S.~K.} \bibnamefont{Ibrahim}},
  \bibinfo{author}{\bibfnamefont{M.}~\bibnamefont{Bruschi}},
  \bibinfo{author}{\bibfnamefont{L.~D.} \bibnamefont{Gioia}},
  \bibinfo{author}{\bibfnamefont{S.~C.} \bibnamefont{Davies}},
  \bibinfo{author}{\bibfnamefont{X.}~\bibnamefont{Yang}},
  \bibinfo{author}{\bibfnamefont{L.-S.} \bibnamefont{Wang}},
  \bibinfo{author}{\bibfnamefont{G.}~\bibnamefont{Sawers}}, \bibnamefont{and}
  \bibinfo{author}{\bibfnamefont{C.~J.} \bibnamefont{Pickett}},
  \bibinfo{journal}{Nature} \textbf{\bibinfo{volume}{433}},
  \bibinfo{pages}{610} (\bibinfo{year}{2005}).

\bibitem[{\citenamefont{Pastawski and Medina}(2001)}]{Pastawski:rmf:01}
\bibinfo{author}{\bibfnamefont{H.~M.} \bibnamefont{Pastawski}}
  \bibnamefont{and} \bibinfo{author}{\bibfnamefont{E.}~\bibnamefont{Medina}},
  \bibinfo{journal}{Revista Mexicana de Fisica} \textbf{\bibinfo{volume}{47}},
  \bibinfo{pages}{1} (\bibinfo{year}{2001}).

\bibitem[{\citenamefont{Hurley et~al.}(1986)\citenamefont{Hurley, Pacios,
  Christiansen, Ross, and Ermler}}]{Hurley:jcp:86}
\bibinfo{author}{\bibfnamefont{M.~M.} \bibnamefont{Hurley}},
  \bibinfo{author}{\bibfnamefont{L.~F.} \bibnamefont{Pacios}},
  \bibinfo{author}{\bibfnamefont{P.~A.} \bibnamefont{Christiansen}},
  \bibinfo{author}{\bibfnamefont{R.~B.} \bibnamefont{Ross}}, \bibnamefont{and}
  \bibinfo{author}{\bibfnamefont{W.~C.} \bibnamefont{Ermler}},
  \bibinfo{journal}{J. Chem. Phys} \textbf{\bibinfo{volume}{84}},
  \bibinfo{pages}{6840} (\bibinfo{year}{1986}).

\bibitem[{\citenamefont{Thygesen}(2006)}]{Thygesen:prb:06}
\bibinfo{author}{\bibfnamefont{K.~S.} \bibnamefont{Thygesen}},
  \bibinfo{journal}{Phys. Rev. B} \textbf{\bibinfo{volume}{73}},
  \bibinfo{pages}{035309} (\bibinfo{year}{2006}).

\end{thebibliography}

\end{document}